\documentclass[12pt]{article}
\pdfoutput=1
\usepackage[totalwidth=480pt,totalheight=640pt]{geometry}
\usepackage[centertags]{amsmath}
\usepackage{array,multirow}
\usepackage{amssymb}
\usepackage{graphicx}
\usepackage{color}
\usepackage{setspace}

\usepackage{epsfig}

\def\Or[#1]{{\text{O}}\left({#1}\right)}
\def\dotl[#1,#2]{\left\langle #1, #2 \right\rangle}
\def\dotlb[#1,#2]{[ #1, #2 ]}
\def\dotp[#1,#2]{(#1) \cdot (#2)}
\def\aff[#1,#2]{\hat{#1}(#2)}
\def\n4sym{{\cal N}=4 SYM}
\def\>{\rangle}
\def\<{\langle}
\def\weight[#1,#2,#3]{\{(#1),#2,#3\}}
\def\ads[#1]{$\text{AdS}_{#1}$}

\newcommand{\ba}{\begin{eqnarray}}
\newcommand{\ea}{\end{eqnarray}}
\newcommand{\be}{\begin{eqnarray}}
\newcommand{\ee}{\end{eqnarray}}

\newcommand{\CA}{{\cal A}}

\newcommand{\CC}{{\cal C}}

\newcommand{\CI}{{\cal I}}
\newcommand{\CL}{{\cal L}}
\newcommand{\CJ}{{\cal J}}

\newcommand{\CN}{{\cal N}}
\newcommand{\CO}{{\cal O}}

\newcommand{\CR}{{\cal R}}
\newcommand{\nn}{\nonumber}

\usepackage{hyperref}

\title{Mellin Amplitudes}

\begin{document}

\begin{titlepage}

\begin{center}
\vspace{1cm}

{\Large \bf  AdS Field Theory from Conformal Field Theory }

\vspace{0.8cm}

\small
\bf{A. Liam Fitzpatrick$^1$,  Jared Kaplan$^{2,3}$}
\normalsize

\vspace{.5cm}

{\it $^1$ Stanford Institute for Theoretical Physics, Stanford University, Stanford, CA 94305}\\
{\it $^2$ SLAC National Accelerator Laboratory, 2575 Sand Hill, Menlo Park, CA 94025} \\
{\it $^3$ Department of Physics and Astronomy, Johns Hopkins University, Baltimore, MD 21218} \\

\end{center}

\vspace{1cm}

\begin{abstract}

We provide necessary and sufficient conditions for a Conformal Field Theory to have a description in terms of a perturbative Effective Field Theory in AdS.  The first two conditions are well-known: the existence of a perturbative `$1/N$' expansion and an approximate Fock space of states generated by a finite number of low-dimension operators.  We add a third condition, that the Mellin amplitudes of the CFT correlators must be well-approximated by functions that are bounded by a polynomial at infinity in Mellin space, or in other words, that the Mellin amplitudes have an effective theory-type expansion.  We explain the relationship between our conditions and unitarity, and provide an analogy with scattering amplitudes that becomes exact in the flat space limit of AdS.  The analysis also yields a simple connection between conformal blocks and AdS diagrams, providing a new calculational tool very much in the spirit of the S-Matrix program.

We also begin to explore the potential pathologies associated with higher spin fields in AdS by generalizing Weinberg's soft theorems to AdS/CFT.  The AdS analog of Weinberg's argument constrains the interactions of conserved currents in CFTs, but there are potential loopholes that are unavailable to theories of massless higher spin particles in flat spacetime.

\end{abstract}

\bigskip

\end{titlepage}

\section{Introduction}
\label{sec:Introduction}

Effective Field Theory may be the most powerful and robust tool available to modern theorists.  Yet only in the last few years has AdS field theory  been systematically examined under the lens of EFT \cite{Katz} in order to improve our understanding of the AdS/CFT correspondence \cite{Maldacena, GKP, Witten}.  And EFT is a lens, for the notion of a cutoff scale provides one of the two clearest and most quantitative metrics for locality.  Scattering provides the other natural test of locality;  it has the advantage of being a completely gauge and reparameterization invariant physical process that can measure the size of objects, detect the presence of long-range forces, and diagnose the breakdown of unitarity. The study of scattering allows one to see how EFTs are restricted by general physical principles, while EFTs in turn make much of the quantitative behavior of scattering amplitudes transparent. Recently there has been great progress connecting CFT correlation functions to scattering processes in AdS \cite{JoaoMellin, Analyticity, Unitarity, Raju:2012zr} using the technology of Mellin space \cite{Mack, Macksummary, NaturalLanguage, Paulos:2011ie, Nandan:2011wc}, the AdS/CFT analog of momentum space.  In this work we will combine insights from EFT and Mellin space with analogies from S-Matrix theory to provide necessary and sufficient criteria for CFTs to have perturbative AdS EFT duals.

In other words, we will give bottom-up criteria for CFTs to have local bulk descriptions.  By `local' we mean that the  dual effective field theory has a cutoff much larger than the AdS curvature scale.  Our perspective is `bottom-up' in terms of the AdS energy scale, or equivalently, the dimension of CFT operators;  we will not have anything to say about the microscopic definition of CFTs in terms of quarks, gluons, or more exotic fundamental constituents.  A major early insight of the AdS/CFT correspondence was that the limit of large radius of curvature of AdS provides a holographic description of scattering amplitudes in flat space through the  correlators of a CFT defined on the boundary \cite{Polchinski, Susskind}.  However, making this connection sharp and transparent has required the development of new techniques, since it is obscured when correlators are written in position space \cite{GGP,TakuyaFSL}.  Since we will be probing bulk locality at distances much smaller than the AdS length, we will also be implicitly studying the question of locality for flat space holography.  Locality in flat spacetime is related to S-Matrix analyticity; throughout we will see deep connections with this subject.

A bottom-up derivation of AdS locality was first attempted using the CFT bootstrap in a seminal paper by Heemskerk, Penedones, Polchinski, and Sully (HPPS) \cite{JP} and was pushed further in subsequent work \cite{Heemskerk:2010ty, ElShowk:2011ag}.  Some aspects of AdS EFT first appear in these works in the guise of anomalous dimensions that grow at large energies in AdS; these results helped to inspire \cite{Katz} and much else.  However, technological restrictions forced HPPS to make very strong assumptions about the behavior of the CFT, essentially ruling out CFT correlators with contributions that would be dual to the exchange of propagating degrees of freedom in AdS.  By formulating the problem in terms of the Mellin amplitude $M(\delta_{ij})$ for CFT correlators we will be able to greatly simplify and generalize the argument.  We will show that if three conditions hold, then the CFT can be viewed as the dual of an AdS field theory:
\begin{description}

\item[\ \ \ 1. Perturbativity] The CFT correlation functions have a perturbative expansion in a parameter that we will call `1/N'.   The CFT is unitary order by order in this parameter  for all states with scaling dimension below a fixed gap $\Delta_\Lambda \gg 1$.
\label{Criteria1}

\item[\ \ \ 2. Fock Space of States] Up to perturbative corrections in $1/N$, the Hilbert space of the CFT for operators of dimension less than $\Delta_\Lambda$ is a Fock space generated by a finite set of primary operators that we will refer to as `single-trace'\footnote{Throughout the paper, our use of the terminology  `$1/N$' and `single-trace' is purely for the sake of familiarity; we will not be making any assumptions about the microscopic physics, or even the spacetime dimension, of the CFT.}. \label{Criteria2}

\item[\ \ \ 3. Polynomial Boundedness] The Mellin amplitudes for the CFT correlation functions are polynomially bounded at large values of the Mellin space variables $\delta_{ab}$, which can be viewed as relative scaling dimensions of CFT operators. 

\end{description} 
The last condition may seem a bit unfamiliar, so let us emphasize that it is implied by a stronger assumption that effective field theorists will recognize:
\begin{description}
\item[\ \  3'. EFT Expansion]   Terms in the Mellin amplitude that grow as $\delta_{ij} \to \infty$ have an expansion in $\frac{\delta_{ij} }{\Delta_\Lambda}$, so the gap $\Delta_\Lambda$ also functions as an EFT-type cutoff.   The Mellin amplitude can be arbitrarily well-approximated for $\delta_{ij}$ below $\Delta_\Lambda$ by keeping only a subset of the terms, after which point the Mellin amplitude will satisfy criterion {\bf 3}.
\end{description}
The first two conditions are fairly well known, but our third criterion is new, and we believe it is of conceptual and practical importance.  Without the criterion {\bf 3} or {\bf 3'} the cutoff of the AdS field theory description could be as low as the AdS curvature scale $1/R$, despite the perturbative expansion in $\frac{1}{N}$.\footnote{The low-energy spectrum of large $N$ confining gauge theories provides an example of this sort of EFT.  The interactions of mesons and glueballs are suppressed by powers of $1/N$, so the first few irrelevant interactions in the EFT for these states would appear to have a cutoff parametrically larger than the confinement scale $\Lambda$.   However, the theory breaks down near $\Lambda$, because general irrelevant operators will be suppressed only by a fixed power of $1/N$ times an arbitrary power of $1/\Lambda$. Here the analogy is between $\Lambda$ and $1/R$ in AdS.}

It will be easy to see that all perturbative AdS field theories give rise to CFT correlation functions that satisfy our criteria. For instance, at tree-level, polynomial boundedness follows straightforwardly by direct computation \cite{JoaoMellin, NaturalLanguage}. Indeed, it is essentially no different from the analogous fact that flat-space effective Lagrangians produce tree-level S-matrices with only propagator poles and polynomial contact interactions in momentum space.  Similarly, effective Lagrangians in AdS produce Mellin amplitudes that are sums over poles and polynomials in the Mellin variables, so that roughly speaking:\footnote{This  counting of derivatives is modified somewhat for interactions including particles with spin. }
\be
\textrm{Interactions in AdS with $2n$ derivatives} &\rightarrow& \textrm{Polynomial of degree $n$ in Mellin variables} , \nn \\
\textrm{Particle Exchange in AdS} &\rightarrow& \textrm{Simple Poles in Mellin Variables} . \nn
\ee

Let us explain the meaning of our three criteria in qualitative terms.  First of all, a perturbative expansion will allow us to sharply distinguish single-trace and multi-trace operators in the CFT.  
This is crucial if we are to make a connection with ``particles'' in the bulk theory, since the notion of single-trace vs. multi-trace is dual to that of single-particle vs. multi-particle states.  
Furthermore, any perturbative bulk dual is by definition an expansion around a free field theory, and free AdS field theories are dual to `infinite $N$' CFTs that are entirely fixed in terms of their spectra. Consequently, the perturbative $1/N$ expansion is important for connecting to perturbation theory in the emergent AdS dual theory.

Knowledge of the spectrum of the CFT determines the locations of all the poles in the Mellin amplitude, providing our most powerful handle on locality.  This follows because singularities in the Mellin amplitude must correspond to the dimensions of operators in the CFT; the residues of these singularities are related to lower-point CFT correlators and their respective Mellin amplitudes through a factorization formula.  Since Mellin amplitudes in unitary CFTs are meromorphic functions (they can have poles but no branch cuts), we can use elementary theorems in complex analysis along with the spectrum of the CFT to determine the Mellin amplitude up to an entire function of the Mellin space variables.  

This also means that there must be an intimate connection between conformal blocks and AdS Feynman diagrams involving particle exchanges, because they must share poles and residues.  In fact, we will see that the difference between the two is simply that the AdS Feynman diagrams are polynomially bounded in Mellin space, while the Conformal Blocks diverge exponentially.  This means that  we can compute AdS Feynman diagrams involving the exchange of spin $\ell$ fields using known formulas for spin $\ell$ conformal blocks \cite{Mack, Macksummary, Analyticity, Unitarity}, without ever invoking an AdS Lagrangian or dealing with issues of higher spin gauge invariance.  This achieves a sort of S-Matrix program for AdS/CFT, because it allows us to compute using only the principles of unitarity, crossing symmetry, and conformal invariance.

Before we explain our last criterion, let us briefly review some basic points about AdS/CFT and the idea of effective conformal field theory \cite{Katz}.  Throughout this paper we will be thinking about CFTs in radial quantization, where operators and states are classified according to their dimension and angular momentum quantum numbers.  CFTs in radial quantization (or in other words, on $R \times S^{d-1}$) are dual to AdS in global coordinates, and the Dilatation operator of the CFT generates time translations in global AdS.  This means that AdS energies are dual to CFT dimensions.  The duality between bulk energies and CFT dimensions implies that effective field theories in AdS with a cutoff $\Lambda$ will be dual to Effective Conformal Theories (ECTs) \cite{Katz} with a cutoff in dimension $\Lambda R$, where $R$ is the AdS scale.  The vast majority of field theories are only EFTs, including all field theories with gravity.  This means that when we ask which CFTs are dual to local AdS field theories, it only makes sense to ask about ECTs and bulk EFTs.

From our point of view, a crucial analogy connects the growth of scattering amplitudes at large energies and the growth of the Mellin amplitude at large dimensions.  If a perturbative and analytic S-Matrix grows no faster than a polynomial at large energies, then it can be parameterized by an effective field theory, where specific monomials in the momenta come from irrelevant operators in the Lagrangian of the EFT.   This EFT description will break down at some cutoff scale where perturbative unitarity bounds are violated, but the EFT will provide an excellent approximation up to that point.  

The analogy with the ECT Mellin amplitude is precise -- as long as it grows only as a polynomial at large $\delta_{ab}$, it can be parameterized by an AdS EFT, which will provide a good description of the CFT physics for operators of dimension less than the ECT cutoff.  There is a perturbative unitarity bound \cite{Katz} on the anomalous dimensions of operators in ECTs, and violations of this bound signal the breakdown of the ECT near the cutoff dimension.  In section \ref{sec:Unitarity} we will explain how CFT perturbative unitarity bounds \cite{Katz} are generically violated if the Mellin amplitude grows too quickly.  This suggests that our condition {\bf 3} can be roughly viewed as a consequence of perturbative unitarity, although for various reasons we will discuss, at this point we cannot give a rigorous proof.

These statements can be summarized by saying that large values of the Mellin space coordinates $\delta_{ab}$ correspond to large dimensions in the CFT, and that this should be thought of as the UV.  CFTs with a perturbation expansion and a Fock space of states that are simple enough in the UV can be described by AdS field theories.

Readers familiar with the S-Matrix program will note that the restrictions we have placed on the UV behavior of the Mellin amplitude are reminiscent of boundedness requirements on the S-Matrix that are traditionally associated with analyticity.  This is no accident.  Following up on the work of Penedones \cite{JoaoMellin}, we showed recently \cite{Analyticity, Unitarity} that the Mellin amplitude provides a holographic definition of the flat space S-Matrix via the flat space limit of AdS/CFT.   We suggested in that work that on a formal level, one can view the analyticity of the S-Matrix, and therefore the fine-grained locality of holographic theories, as a consequence of the extremely constrained analytic structure of the Mellin amplitude.  Our polynomial boundedness criterion makes these ideas precise in the case where `locality' means the existence of an AdS effective field theory.

Since we are discussing the holographic emergence of local EFT, we would like to mention an analogy between Weinberg's perspective \cite{Weinberg:1995mt} on flat space quantum field theory and a similar perspective on AdS \cite{Katz}.  Weinberg has argued that if one wants to construct a perturbative S-Matrix satisfying the principles of quantum mechanics, Lorentz invariance, and cluster decomposition with a finite number of particle species, then one will be led inexorably to local quantum field theory in flat spacetime.  Similarly, we would argue that to construct perturbative CFT correlation functions for a finite number of operator species, we are naturally led to the study of local AdS field theory. 
In both the S-Matrix and CFT case this viewpoint `derives' a higher dimensional spacetime as a necessary holographic arena. 

So far we have ignored the question of spin.  There are well-known difficulties in formulating consistent interacting field theories involving particles with spin $> 2$, especially in situations where the background curvature is negligible (e.g.~in flat spacetime).  If the spectrum of a CFT contains single-trace higher spin currents, then its AdS dual must contain corresponding higher spin fields, so these difficulties must also be visible from the point of view of the CFT and its correlation functions.  And in fact we have evidence that CFTs satisfying our hypotheses, such as N=4 SYM in the large $N$ and large 't Hooft coupling limit, do not have low dimension single-trace operators with spin $>2$.  Therefore one can ask if it is possible to understand these facts from the bottom-up, directly in terms of CFT correlation functions.

We can understand the limitations on low-dimension higher spin currents by studying a particular kinematic limit of CFT correlators in the Mellin representation.  This kinematical limit simply represents the soft emission of higher spin fields in AdS.  In the case of the flat space S-Matrix, it was shown long ago \cite{Weinberg:1965nx} that by studying the soft emission of higher spin massless particles, one can prove that these particles cannot give rise to any interactions that survive at long distances and lead to long-range forces (for a review and references see \cite{Porrati:2008rm}).  We will see that equivalent results can be obtained for AdS/CFT correlators, with the Mellin-space calculations displaying a striking technical similarity to the older computations in momentum space.  However, unlike the case of flat spacetime, the arguments are only approximate when the cutoff $\Lambda R$ on CFT dimensions is finite, and so there are non-trivial loopholes which can be understood as a consequence of the intrinsic IR cutoff in the AdS `cavity'.  It would be interesting to understand how known theories \cite{Vasiliev:1990en, Vasiliev1,Vasiliev2} of higher spin fields in AdS fit through these loopholes.

The outline of this paper is as follows.  In section \ref{sec:MellinReview} and \ref{sec:MellinPoles} we review some basic features of AdS/CFT and Mellin space, emphasizing what sort of Mellin amplitudes arise from AdS field theories and the crucial connection between poles in the Mellin amplitude, the CFT spectrum, and factorization.  In section \ref{sec:ConformalBlocksandDiagrams} we explain the connection between conformal blocks and AdS exchange diagrams, and show how this can be used as a concrete computational tool that avoids the complexities of AdS Lagrangians for fields of arbitrary spin.  With these tools in hand, we explain the general argument leading from our CFT criteria to AdS field theory in section \ref{sec:AdSFieldTheoryPolynomialBoundedness}.  In section \ref{sec:Unitarity} we explain the relationship between unitarity and our criterion, emphasizing how the growth of the Mellin amplitude can lead to  violations of perturbative unitarity \cite{Katz} in analogy with the growth of the S-Matrix in the UV.  In section \ref{sec:FSLandAnalyticity} we explain the analogies and the connections between our argument and more familiar methods in S-Matrix theory; this section may be particularly useful for readers comfortable with scattering amplitudes but less familiar with CFTs and Mellin amplitudes.  Finally in section \ref{sec:HigherSpin} we explain how to generalize Weinberg's soft theorems from scattering theory to AdS/CFT, first with a review and a simple example, and then with a somewhat more general treatment of currents with arbitrary spin $\ell$.  We conclude and discuss the results in section \ref{sec:Discussion}. In the appendices, we include the details of a calculation necessary for section \ref{sec:HigherSpin} and discuss an alternate approach to AdS locality by Sundrum \cite{Sundrum:2011ic}.

\section{AdS Locality from Polynomial Boundedness}

\subsection{Mellin Amplitudes for AdS Field Theory}
\label{sec:MellinReview}

Let us begin by reviewing Mellin amplitudes for the boundary correlators of AdS effective field theories, in order to see why they automatically satisfy the three criteria discussed in the introduction.  Since we are starting with an AdS EFT, the energy gap $\Delta_\Lambda R^{-1}$ is also the cutoff. We will begin by restricting to tree-level amplitudes of scalars, and then discuss how their behavior extends to spin $\ell$ operators and loop corrections.

  First, recall the definition of the Mellin amplitude.  In the case of correlators of scalar CFT operators, it takes the form\footnote{For more details of the integration measure and contour, see Appendix A of \cite{JoaoMellin} or \cite{Analyticity}.  We will mostly be analyzing 4-pt correlators, in which case the Mellin amplitude is unique and the Mellin transform is invertible; for discussions of uniqueness in the case of general $n$-pt correlators see Appendix A of \cite{Analyticity}.}
\be
 \< \CO_1(P_1) \CO_2(P_2) \dots \CO_n(P_n) \> &=&  \int [d \delta_{ij}] M(\delta_{ij}) \prod_{i<j}^n \Gamma(\delta_{ij}) P_{ij}^{-\delta_{ij}} ,
\label{eq:mellindef}
\ee
where $P_{ij} = -2 P_i \cdot P_j$ with $P_i^A $ being vectors in the $(d+2)$-dimensional embedding space \cite{Ferrara:1973yt,Dirac:1936fq,Mack:1969rr,Ferrara:1973eg,ourDIS,DSDProjectors, Costa2011mg,Weinberg:2010fx, SimmonsDuffin:2012uy}. These satisfy $P^2=0$ and are identified under $P^A \sim \lambda P^A$ for non-zero $\lambda$;  by choosing $\lambda$ so that $P^+ = 1$ the $P^A$ can be related to the usual coordinates in flat Euclidean space by 
\be
P^A=(P^+,P^-,P^\mu) = (1, x^2, x^\mu).
\ee
  The embedding space metric is 
  \be
  ds^2 = -dP^+ dP^- + dP^\mu dP_\mu, 
  \ee
so $P_{ij} =(x_i - x_j)^2$.  
For a nice review of the embedding space formalism in conformal field theory, see Section 2 of \cite{Costa2011mg}. To extend the embedding space formalism to AdS, one simply considers points $X$ on the hyperboloid $X^2 = -1$ \cite{ourDIS,JoaoMellin}; these points are not identified under a rescaling.  

  Not all of the $\delta_{ij}$'s in the integration measure of equation (\ref{eq:mellindef}) are independent; they are symmetric ($\delta_{ij} = \delta_{ji}$), have zero's on the diagonal $(\delta_{ii} = 0)$, and satisfy constraints related to the dimensions $\Delta_i$ of the external operators $\CO_i$:
\be
\sum_{j \ne i } \delta_{ij} = \Delta_i .
\label{eq:deltaConstraint}
\ee
These constraints are identical in form to those obeyed by the Mandelstam invariants $s_{ij} = p_i \cdot p_j$ of an $n$-particle scattering amplitude, and in the case of $n=4$, there are only $2$ independent $\delta_{ij}$.  Correlators of symmetric traceless tensor operators $\CO^{A_1 \dots A_\ell}$ can then easily be included by associating an additional ``polarization'' vector $Z^A$ with each operator \cite{Costa2011mg}, and taking
  \be
  \CO_i(P_i, Z_i) \equiv Z^{A_1}_i \dots Z^{A_\ell}_i \CO_{A_1 \dots A_\ell}(P_i).
  \ee
The Mellin amplitude can then still defined as in equation (\ref{eq:mellindef}), with $n$ additional ``points'' $P_{n+i} \equiv Z_i$ and external ``dimensions'' $\Delta_{n+i} \equiv -\ell_i$ in the constraints equation (\ref{eq:deltaConstraint}).  
  Furthermore, formally, the $Z_i$ vectors satisfy $Z_i \cdot P_i = 0$ and $  Z_i^2=0$ in order to represent symmetric traceless tensors, as explained in \cite{Costa2011mg}.  One can also expand in $Z_i \cdot P_j$ and compute a number of separate Mellin amplitudes corresponding to each polarization structure, in the same way that one obtains a different scattering amplitude for each combination of external helicities.

Next, we recall that the $n$-point Mellin amplitude from a simple scalar contact interaction $\CL_{\rm AdS} = g \phi^n$ is just the coupling times a normalization constant,\footnote{Explicitly, the normalization is
\be
\lambda_n= \frac{\pi^{\frac{d}{2}}}{2} \Gamma\left(   \frac{1}{2} \sum_{i=1}^n \Delta_i - \frac{d}{2}\right) \prod_{i=1}^n \CC_{\Delta_i} ,
\label{eq:lambdan}
 \ee
with $\CC_\Delta \equiv \frac{\Gamma(\Delta)}{2\pi^{d/2} \Gamma(\Delta - \frac{d}{2} +1)}$. } independent of $\delta_{ij}$:
\be
M_n(\delta_{ij}) = g \lambda_n .
\label{eq:mellincontact}
\ee
 In fact, one of the crucial simplifications of correlators in Mellin space is that all local interactions in AdS produce polynomial Mellin amplitudes at tree-level, as follows. Take an AdS Lagrangian with the contact interaction
\be
S_{\rm AdS} = \int_{\rm AdS} d^{d+1} X \prod_{i, j} (\nabla_i \cdot \nabla_j)^{a_{ij}} \phi_1 ... \phi_n,
\ee
where $\nabla_i$ is notation for an AdS covariant derivative on $\phi_i$ and we have written the integration over AdS space $d^{d+1} X$ using embedding coordinates.  An $n$-point correlator following from this interaction can be calculated from
\be
A(P_i) = \int_{\rm AdS} d^{d+1} X  \prod_{i,j = 1}^n (\nabla_i \cdot \nabla_j)^{a_{ij}} \prod_{i=1}^n G_{\partial B} (P_i, X),
\ee
where the derivative $\nabla_i$ acts on the bulk-to-boundary propagator $G_{\partial B} (P_i, X) = \frac{\CC_\Delta}{(-2P_i \cdot X)^\Delta}$ and differentiates it with respect to $X$ along the AdS hyperbola \cite{GGP, JP}.  At the level of the Mellin amplitude, derivatives can be added recursively using
\be
(\nabla_i \cdot \nabla_j ) M_{\Delta_a}(\delta_{ab}) = \left( \Delta_i \Delta_j  - 2 \delta_{ij} \right) M_{\Delta_a}(\delta_{ab}) .
\ee
This relation also makes it clear that we can obtain every polynomial $P(\delta_{ij})$ in Mellin space using a linear combination of AdS derivative interactions.  Similar results have been obtained for particles with more general spin $\ell$ in \cite{Paulos:2011ie}.

 Mellin amplitudes for particle exchanges in AdS are also very simple. As shown in \cite{NaturalLanguage}, AdS Feynman diagrams for tree-level exchange of particles in AdS factorize in Mellin space on a tower of poles associated with the primary and descendant operators dual to the exchanged particle.  This factorization formula takes the form
\be
M(\delta_{ij})|_{\rm poles} =-4 \pi^{\frac{d}{2}}  \sum_{m=0}^\infty \left( \frac{\Gamma(\Delta-\frac{d}{2}+1)  m!}{(\Delta-\frac{d}{2}+1)_m}\right) \frac{ \,\CL_m(\delta_{ij})\,\CR_m(\delta_{ij})}{\delta- (\Delta+2m)},
\label{MainFormula}
\ee
where $\delta$ is the following linear combination of $\delta_{ij}$'s:
\be
  \delta = \sum_{i=1}^k \Delta_i - 2\sum_{i < j \leq k} \delta_{ij},
\ee
and the residue factors $\CL_m, \CR_m$ are fixed by the Mellin amplitudes on the left and right of the diagram, obtained by cutting the propagator that is being factorized \cite{NaturalLanguage}.

Thus, by direct calculation, one sees that AdS effective field theories produce tree-level Mellin amplitudes that are bounded by polynomials with a maximum degree determined by the highest dimension interaction in the effective Lagrangian. This applies to effective theories in the sense that one always works to some order of accuracy at low energies, and consequently one can always truncate the effective Lagrangian to a finite number of interactions determined by the desired level of precision. Of course, one may prefer to think of effective theories as having an infinite number of operators of increasingly large dimensions, suppressed by increasing powers of the cutoff. In this case, it remains obvious that the Mellin amplitude will satisfy our weaker criteria {\bf 3'} -- it will have an EFT expansion.  Polynomial boundedness will then obtain to any finite order in inverse powers of the cutoff $\Delta_\Lambda R^{-1}$. 

In fact, the ability to take into account an infinite number of higher-dimensional interactions is necessary if we want to include loop corrections in general, since for non-renormalizable theories they will be generated radiatively in any case even if they are not present at tree-level.  However, at any finite order in $1/\Delta_\Lambda$, there will be a maximum power with which a Mellin amplitude can grow, since in the limit of $\delta_{ij} \gg 1$, Mellin amplitudes behave the same way in the $\delta_{ij}$'s as flat-space scattering amplitudes behave in Mandelstam invariants.\footnote{We should note that the current state of computing loop diagrams for Mellin amplitudes is not yet at a point where this is technically straightforward to demonstrate by direct computation.  Loop integrals for a limited set of diagrams can be simplified to sums over tree-level diagrams by essentially using a K\"allen-Lehman-esque representation \cite{Analyticity}, but general loop diagrams are quite complicated and better computational techniques are required.  The most transparent argument for why loop diagrams should grow in their Mellin variables according to the same power-counting rules as for flat-space amplitudes in their Mandelstam invariants is that a simple integral transform relates the large $\delta_{ij}$ of the former to the latter \cite{JoaoMellin,NaturalLanguage,Analyticity}.  Furthermore, this transform and its inverse convert monomials into monomials.  }  In any case, once we demonstrate the existence of an AdS EFT at leading order in $1/N$, the UV insensitive higher order corrections will match between the CFT and the AdS field theory in order to simultaneously satisfy the unitarity relations in both AdS and the CFT \cite{Unitarity}.

\subsection{CFT States and Mellin Poles}
\label{sec:MellinPoles}

The large $N$ limit provides an important tool for analyzing CFT behavior.  One often constructs large $N$ CFTs using non-abelian gauge fields with rank $N$, but more generally, by the large $N$ limit we  mean a limit where CFT correlators approximately factorize into products of two-point functions of special, `single-trace' operators.  More precisely, there exists a set of ``primary single-trace'' operators $\{ \Phi_a \}$ that at leading order in $1/N$ have only disconnected correlators:
\be
\< \Phi_1 (x_1) \Phi_2 (x_2) \dots \Phi_{2n}(x_{2n}) \> &=& \< \Phi_1(x_1) \Phi_2(x_2) \> \dots  \< \Phi_{2n-1} (x_{2n-1}) \Phi_{2n} (x_{2n})\> + \textrm{permutations} \nn\\
 && + \textrm{ suppressed by } 1/N.
 \label{eq:disconnected}
\ee
Furthermore, all local operators in the theory can be constructed by taking products and derivatives of the  $\Phi_a$'s. A concise way of stating this assumption is that as $N\rightarrow \infty$, the generating functional of the correlators of the theory is simply a Gaussian in the $\Phi_a$'s.  Crucially, the dynamics at large $N$ are simply those of uncoupled harmonic oscillators.

Here, we will briefly review these phenomena; more thorough reviews are given in \cite{JP,JPStringTheory,Ginsparg}. The states of the theory are obtained by radially quantizing the CFT, so that each local operator $\CO(x)$ corresponds to a state through its action on the vacuum at the origin:
\be
| \CO \> \equiv \CO(0) |0 \> .
\ee
The role of the Hamiltonian in radial quantization is played by the dilatation operator $D$, which generates radial rescalings (i.e. dilatations), and its eigenstates are given by the operators of definite dimension:
\be
D |\CO\> = \Delta_\CO | \CO \>.
\ee
It is convenient to introduce a creation operator  and annihilation operator, $a^\dagger_{\Phi_a}$ and $a_{\Phi_a}$, for all the primary single-trace operators $\Phi_a$ in the theory. Derivatives of $\Phi_a$'s give additional operators with dimensions related by integers
\be
D | \partial^n \Phi \> = (\Delta_\Phi + n) | \partial^n \Phi \>.
\ee
Such states are called ``descendants'', and they get their own creation and annihilation operators $a^\dagger_{\partial^n \Phi_a}, a_{\partial^n \Phi_a}$.  Together with their corresponding primary operator, these descendants are needed in order to fill out complete irreducible representations of the conformal algebra, called conformal blocks.  This can be seen explicitly by the fact that the generator $P^\mu$ of translations acts on local operators through derivatives:
\be
\left[ P_\mu, \Phi(x) \right] = i \partial_\mu \Phi(x).
\ee

The reason for introducing the  creation and annihilation operators is that all remaining states in the theory are products of the single-traces, and thus the space of states  is a Fock space.  Furthermore, the dimension of products of operators is just given by sums of the individual dimensions, up to $1/N$ corrections:
\be
\Delta_{\Phi_1 \Phi_2} = \Delta_{\Phi_1} + \Delta_{\Phi_2} + \CO(1/N).
\label{eq:infiniteNdims}
\ee
Therefore, the infinite $N$ limit of the dilatation operator takes the form of the Hamiltonian for a simple harmonic oscillator:
\be
D &=& \sum_i \Delta_i a^\dagger_i a_i.
\ee

An essential tool for understanding conformal theories is the operator product expansion (OPE), which allows one to express products of local operators as sums of operators times spacetime-dependent coefficients:
\be
\CO_1(x) \CO_2(0) &=& \sum_\CO c_{\CO}(x)^{\mu_1 \dots \mu_\ell} \CO_{\mu_1 \dots \mu_\ell}(0).
\label{eq:OPE}
\ee
At infinite $N$, the only operators that appear on the RHS of equation (\ref{eq:OPE}) are double-trace operators (states created by exactly two creation operators).
This can be seen explicitly by decomposing the infinite $N$ correlators in equation (\ref{eq:disconnected}) into the contributions from individual conformal blocks.  For concreteness, consider the example of the four-point function $\< \CO_1 \CO_2 \CO_1 \CO_2\>$. This can be decomposed as
\be
&&\< \CO_1(x_1) \CO_2(x_2) \CO_1(x_3) \CO_2(x_4) \> = \frac{\CC_{\Delta_1}\CC_{\Delta_2}}{x_{13}^{2\Delta_1} x_{24}^{2\Delta_2}} = \left( \frac{x_{24}}{x_{13}} \right)^{\Delta_1 - \Delta_2} \sum_{n,\ell} c_{n,\ell}^2 \frac{g_{\Delta_{n,\ell}, \ell}(u,v)}{x_{12}^{\Delta_1 + \Delta_2} x_{34}^{\Delta_1 + \Delta_2}}, 
\ee
where $\CC_\Delta$ 
is a conventional normalization factor, $c_{n,\ell}$ are the OPE coefficients of the conformal blocks, and $g_{\Delta,\ell}(u,v)$ are functions of the conformal invariant cross-ratios 
\be
u= \left( \frac{x_{12}^2 x_{34}^2}{x_{13}^2 x_{24}^2 } \right),
v= \left( \frac{x_{14}^2 x_{23}^2}{x_{13}^2 x_{24}^2 } \right).
\ee
The functional form of the conformal block functions $g_{\Delta,\ell}(u,v)$ (often just called `conformal blocks') only depends on the dimension $\Delta$ and spin $\ell$ of the blocks themselves \cite{Dolan:2000ut}, along with the spacetime dimension. At infinite $N$ there is an infinite tower of double-trace operators for each integer $\ell$, with dimensions from equation (\ref{eq:infiniteNdims})
\be
\Delta_{n,\ell} = \Delta_1 + \Delta_2 +2 n + \ell,
\label{eq:MeanFieldDims}
\ee
and OPE coefficients that were computed in \cite{Unitarity}
\be
&&c_{n\ell}^2    = \frac{ \CC_{\Delta_1} \CC_{\Delta_2}  (-1)^\ell (\Delta_1-\frac{d}{2}+1)_n (\Delta_2-\frac{d}{2}+1)_n
   (\Delta_1)_{\ell+n} (\Delta_2)_{\ell+n}}{\ell! n! (\ell+\frac{d}{2})_n (\Delta_1+\Delta_2+n-2\frac{d}{2}+1)_n (\Delta_1+\Delta_2+2 n+\ell-1)_l (\Delta_1+\Delta_2+n+\ell-\frac{d}{2})_n} ,\nn\\
\ee
where $(a)_m\equiv \frac{\Gamma(a+m)}{\Gamma(a)}$ is the Pochhammer symbol.

  At subleading order in $1/N$, three effects occur.  First of all, the OPE coefficients of the double-trace operators get corrections.  More interestingly, new operators appear on the RHS that were not present at infinite $N$, with OPE coefficients that are $\CO(1/N)$ or higher.  Finally, the dimensions of operators shift by $\CO(1/N)$ anomalous dimensions.

A key property of Mellin amplitudes is that they are meromorphic, with simple poles corresponding to the presence of operators in the OPE.   For example, 
 in the case of a four-point function of four different operators
\be
\CA(x_i) &=& \< \CO_1(x_1) \CO_2(x_2) \CO_3(x_3) \CO_4(x_4) \>,
\ee
there is no disconnected contribution, so the correlator vanishes at infinite $N$.  At subleading order, there may be operators that appear in the OPE of both, say, $\CO_1(x) \CO_2(0)$ and $\CO_3(x) \CO_4(0)$:
\be
\CO_1(x) \CO_2(0) = c_{12\Phi}(x) \Phi(0) + \dots, \nn\\
\CO_3(x) \CO_4(0) = c_{34 \Phi}(x) \Phi(0) + \dots,
\ee
in which case the Mellin amplitude for $\CA(x_i)$ contains a pole at the twist $\tau_\Phi = \Delta_\Phi - \ell_\Phi$ of $\Phi$:
\be
M(\delta_{ij}) = \frac{f(\delta_{ij})}{\delta - \tau_\Phi} + \dots,
\ee
where $\delta = \Delta_1 + \Delta_2 - 2 \delta_{12}$.  However, as we have seen, a primary operator $\Phi$ is always accompanied by its descendants, 
so a full conformal block will contribute a whole tower of poles with relative residues fixed by conformal symmetry:
\be
M(\delta_{ij}) = \sum_m \frac{R_m(\delta_{ij})}{\delta - \tau_\Phi -2m} + \dots,
\ee
where the $\dots$ stand not only for any other operators that occur in the OPE but also for non-pole contributions from the conformal block.   

There is an important distinction to be made for poles we described above as compared to the poles corresponding to double-trace operators, because 
double-trace poles are factored out of the Mellin amplitude by hand in order to separate out kinematic from dynamic information.  More precisely, the integrand of the Mellin transform in equation (\ref{eq:mellindef}) contains not just the Mellin amplitude $M(\delta_{ij})$ itself, but also $\Gamma$ functions:
\be
\prod_{i<j} \Gamma(\delta_{ij}).
\ee
The relevant point is that there are poles that do not occur in  $M(\delta_{ij})$, but do occur in the integrand.  These poles correspond to the multi-trace operators that are present in the OPE even in the infinite $N$ limit. Probably the clearest way to think about these poles is through examples of tree-level contact interactions in AdS.  Specifically, let us return to the contact interaction $g \phi_1^2 \phi_2^2 $ which is of the form discussed surrounding equation (\ref{eq:mellincontact}):
\be
M_n(\delta_{ij}) = g \lambda_n.
\ee  
This Mellin amplitude is simply a constant, exactly analogous to flat-space tree-level amplitudes for contact interactions in QFT. Since the Mellin amplitude itself contains no poles, the only conformal blocks associated with this correlator are those with poles coming from the $\Gamma(\delta_{ij})$ functions.  To see explicitly that these correspond with double-trace operators, first use the constraints on the $\delta_{ij}$ variables to obtain
\begin{equation}
\CA = g \lambda_n \left( \frac{ x_{24}}{x_{13}} \right)^{\Delta_1 - \Delta_2}\int_{-i \infty}^{i \infty} \frac{d \delta_{12} d \delta_{14}}{(2\pi i )^2} \Gamma^2( \delta_{12}) \Gamma^2(\delta_{14}) \Gamma(\Delta_1 - \delta_{12} - \delta_{14}) \Gamma(\Delta_2 - \delta_{12} - \delta_{14}) \left(  u^{\frac{\Delta_1 + \Delta_2}{2} - \delta_{12}} v^{-\delta_{14}}  \right) .
\end{equation}
Closing the $\delta_{12}$ in the left half-plane, we pick up the residues of a tower of poles at 
\be
\delta_{12} = 0,-1,-2, \dots ,
\ee
and the corresponding powers of $u$ in the residue are of the form $\frac{\Delta_1 + \Delta_2}{2} + m$, corresponding to blocks with dimensions of the form\footnote{For blocks of primaries with odd spin, the power of $v$ can contribute odd integers to the dimension. }  \cite{Dolan:2000ut}
\be
\Delta_1 + \Delta_2 + 2n+\ell,
\ee  
in agreement with the dimensions of double-trace operators in equation (\ref{eq:MeanFieldDims}).  The presence of double-poles implies that there is also a residue term that differentiates $\delta_{12}$ in the integrand, thus producing $\log(u)$ terms.  These correspond to anomalous dimensions that are generated by the AdS contact interaction \cite{Dolan:2000ut, Freedman:1998bj,LiuTseytlin,Fitzpatrick:2011hh}.

\subsection{Conformal Blocks and AdS Feynman Diagrams}
\label{sec:ConformalBlocksandDiagrams}

Conformal blocks are the natural building blocks for CFT correlation functions, playing the same role in CFT correlators that partial waves play in scattering amplitudes.  A position-space conformal block for a CFT 4-pt correlator is a function of the conformally invariant cross ratios $u,v$ corresponding to the exchange of an irreducible representation of the conformal group between two pairs of operators.  One way to derive these functions is by using the OPE to express a CFT 4-pt correlator as a sum over 2-pt correlators multiplied by OPE coefficients, as we discussed above.  An easier derivation, which makes the analogy with scattering more immediate, is to insert a sum over states into the CFT correlator as
\be
B_{\Delta, \ell}(x_i)  = \left\langle \CO_1(x_1) \CO_2(x_2) \left( \sum_{\alpha  =  {\rm prim, desc } \ \! \Delta, \ell} | \alpha \rangle \langle \alpha | \right) \CO_3(x_3) \CO_4(x_4) \right\rangle 
\ee
and organize it according to irreducible representations of the conformal group.  By the operator-state correspondence this can be re-written as an integral over a product of CFT 3-pt functions, as discussed recently in \cite{DSDProjectors}.  

\begin{figure}
\begin{center}
\includegraphics[width=\textwidth]{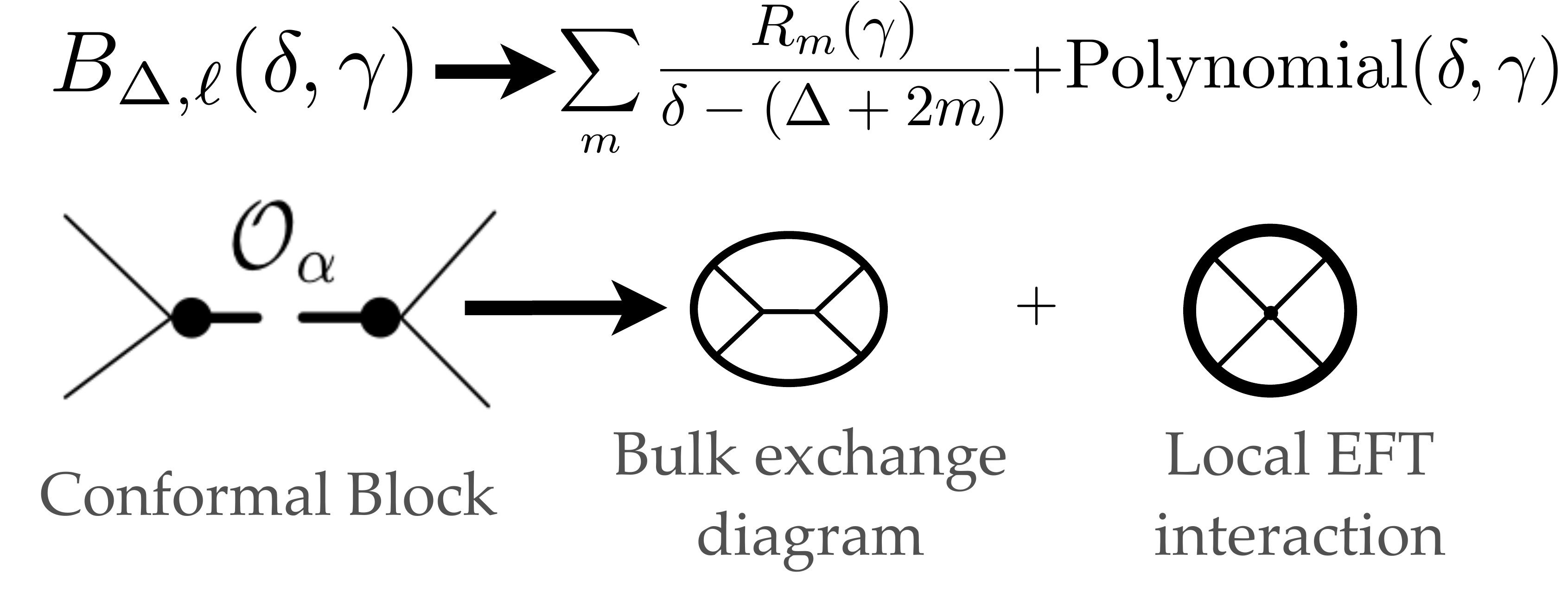}
\caption{This figure shows what happens when one drops the exponentially growing part of the Mellin amplitude for a spin $\ell$ conformal block.  The block turns into an AdS exchange Feynman diagram for a spin $\ell$ particle plus AdS contact interactions. }
\end{center}
\end{figure}

In Mellin space, the contribution to a four-point function of scalars $\CO_i$ from a conformal block of dimension $\Delta$ and spin $\ell$ takes the form \cite{Mack, Analyticity}
\be
\label{eq:ConformalBlock}
B_\Delta^\ell(\delta_{ij}) = e^{ \pi i(h- \Delta)} \left( e^{i \pi (\delta+ \Delta -2h)} -1\right)  \frac{\Gamma \left(\frac{\Delta-\ell-\delta}{2} \right) \Gamma \left(\frac{2h-\Delta - \ell - \delta}{2} \right) }{ \Gamma \left(\Delta_a-\frac{\delta}{2} \right) \
 \Gamma \left(\Delta_b - \frac{\delta}{2} \right) } P_{\ell}(\delta_{ij}) ,
\ee
where $P_\ell(\delta_{ij})$ is a degree $\ell$ ``Mack'' polynomial \cite{Mack, Analyticity} in the $\delta_{ij}$'s, and $2 \Delta_a = \Delta_1 + \Delta_2$ while $2\Delta_b = \Delta_3 + \Delta_4$, with $\Delta_i$ being the dimension of $\CO_i$.  The gamma functions in the numerator have poles at $\delta = \Delta - \ell + 2n$ that correspond to the primary and descendant operators of the conformal block.  However, there are also ``shadow'' poles at $\delta = 2h - \Delta-\ell +2n$ that do not correspond to any physical states.\footnote{For $\Delta > h+1$, this is especially clear, since the primary state would have dimension $2h-\Delta-\ell$, which is below the unitarity bound.}  This necessitates the exponential prefactor, which has zeros that cancel these unphysical poles.  

There are directions in the complex plane where the conformal block blows up exponentially.  We saw above that CFT correlators derived from AdS Field Theories always produce Mellin amplitudes that are bounded by polynomials at large $\delta_{ij}$.  This immediately implies that a lone conformal block cannot arise from AdS Field Theory\footnote{This is no surprise, because only infinite sums of conformal blocks can satisfy both crossing symmetry and unitarity \cite{JP}.}.   But there is an extremely simple operation that we can perform on the block to eliminate this problem -- we can express it as a sum over its poles, and then simply discard the rest.  This procedure transforms the block in a very physical way, because it leaves the poles and residues of the block, which correspond to the exchange of specific states, completely intact.

As a concrete example, let us consider the scalar conformal block
\be
 \label{eq:ConformalBlock}
B_\Delta^0(\delta_{ij}) = e^{ \pi i(h- \Delta)} \left( e^{i \pi (\delta+ \Delta -2h)} -1\right)  \frac{\Gamma \left(\frac{\Delta-\delta}{2} \right) \Gamma \left(\frac{2h-\Delta - \delta}{2} \right) }{ \Gamma \left(\Delta_a-\frac{\delta}{2} \right) \
 \Gamma \left(\Delta_b - \frac{\delta}{2} \right) } .
\ee
This function has poles at $\delta = \Delta + 2m$ for integers $m$, so keeping only the residues of these poles transforms the block into the function 
\be
M_\Delta(\delta) = \sum_{m=0}^\infty \left( \frac{  2i \sin(\pi (\Delta - h) )  \Gamma (h - \Delta - m)}{m! \Gamma \left( \Delta_a - \frac{\Delta}{2} - m \right) \Gamma \left( \Delta_b - \frac{\Delta}{2} - m \right) } \right) \frac{1}{\delta - (\Delta + 2m)} .
\ee
The sum converges for all values of $\Delta$ as long as the spacetime dimension is positive.  Up to a normalization factor, this is precisely the form of the Mellin amplitude for an AdS Feynman diagram corresponding to a bulk scalar exchange.  Thus we see that up to a polynomial indicative of 4-pt contact interactions, the unique polynomially bounded Mellin amplitude including a scalar conformal block corresponds to scalar exchange in AdS!

As another example let us consider graviton exchange between massless scalars.  The conformal block for the exchange of the energy momentum tensor $T_{\mu \nu}$ in dimension $d = 2h$ takes the form
\be
B^{\ell = 2}_{\Delta = d} = \left(\gamma ^2 (4-8 h)+\delta  (\delta -4 h+4)\right)
\frac{\left(-1+e^{i \pi  \delta }\right) \Gamma \left(-\frac{\delta
   }{2}-1\right) 
   \Gamma \left(h-\frac{\delta }{2}-1\right)}{4 \Gamma^2 \left(2 h-\frac{\delta
   }{2}\right)}.
\ee
This has poles when $\delta = 2h-2, 2h, 2h+2,...,4h$, but at large $\delta$ it blows up exponentially due to the phase factor.  If write it as a sum over poles and drop all other contributions, we find
\be
\sum_{m=0}^{h}  \frac{1}{m! (h+m)! \Gamma^2(h+1-m) }  \left( \frac{ \gamma ^2-h \left(2 \gamma ^2+h-2\right)+m^2-1 }{\delta - 2(h - 1 + m)} \right).
\ee
This matches the results of \cite{JoaoMellin} for a scalar minimally coupled to AdS$_5$ gravity,
\be
M_{{\rm AdS}_5  } \propto  \frac{6 \gamma ^2+2}{\delta -2} + \frac{8 \gamma ^2}{\delta -4}+\frac{\gamma ^2-1}{\delta
   -6} - \frac{15}{3} \delta + \frac{55}{2},
\ee
up to the last two terms, which are polynomials that cannot be extracted from the pole structure.   These polynomial terms can be reproduced, or cancelled, by including scalar contact interactions as discussed in section \ref{sec:MellinReview}.  Using the technology of \cite{Costa2011mg, Costa:2011dw} one could use this method to quickly compute the AdS Feynman diagrams for graviton exchange between gravitons, without ever making reference to a bulk Lagrangian.  For fields with general spin $\ell$ this provides an extremely simple way to calculate while avoiding all the complexities of higher-spin gauge invariance.

In summary, imposing Mellin space polynomial boundedness on conformal blocks without compromising their pole and residue structure necessarily turns them into Mellin amplitudes for AdS exchange diagrams plus possible polynomial contact terms.  We are now ready to argue that CFTs obeying our three criteria have a description in terms of AdS effective field theory.

\subsection{AdS Field Theory from Polynomial Boundedness}
\label{sec:AdSFieldTheoryPolynomialBoundedness}

In the previous sections we assembled the ingredients we need to classify the CFTs that can be described by effective field theories in AdS.  Given a CFT that satisfies our first two criteria,
we can list the `single-trace' primary operators $\CO_i$ and their non-vanishing 3-pt functions
\be
\label{eq:3ptOPECoefs}
\langle \CO_i^{\mu_1... \mu_{s_i}} \CO_j^{\nu_1... \nu_{s_j}} \CO_k^{\sigma_1... \sigma_{s_k}} \rangle = \frac{1}{N} \sum_{a} C_{ijk}^a F_{a}^{\mu_i \nu_j \sigma_k }(x_i, x_j, x_k).
\ee
The functions $F_a$ encode the universal tensor structures available for 3-pt functions of primary operators of spin $s_i, s_j,$ and $s_k$, while the $C_{ijk}^a$ are their coefficients.  Only the $C_{ijk}^a$ are dynamical information about the CFT, since the $F_a$ are fixed by symmetry.  We have included an overall factor of $1/N$ to emphasize the fact that we are working in perturbation theory around the Fock space generated by the primary operators $\CO_i$.

Now let us consider the 4-pt correlation functions of these single-trace operators,
\be
\langle \CO_i \CO_j \CO_k \CO_l \rangle = \int [d \delta] M_{ijkl}(\delta_{ab}) \prod_{a < b}^4 \Gamma(\delta_{ab}) P_{ab}^{-\delta_{ab}},
\ee
expressed in terms of their Mellin amplitudes.  As explained in section \ref{sec:MellinPoles}, these Mellin amplitudes can only have poles that correspond with CFT states/operators.  Using our criteria {\bf 1} on the perturbativity of the CFT spectrum, this means that every pole in the Mellin amplitude must come from some particular primary single or multi-trace operator.  

Poles in the Mellin amplitude\footnote{It is crucial here that we are discussing the Mellin {\it amplitude}, and not the Mellin {\it integrand}.  The integrand has $\Gamma(\delta_{ij})$ functions that produce poles associated with multi-trace operators, but these purely `kinematic' poles combine infinite $N$ OPE coefficients with the interacting coefficients, as discussed above and in \cite{Unitarity}.  }  from multi-trace operators necessarily arise from interactions beyond the leading order in perturbation theory.  This follows because the residues of these poles are proportional to the product of two interacting (order $1/N$ or higher) OPE coefficients.  Each of these $O \! \left( \frac{1}{N} \right)$ OPE coefficients are proportional to the 3-pt correlator of two single-trace operators and the intermediate $k$-trace operator.  Furthermore, every $k$-trace operator has an OPE coefficient at zeroth order in perturbation theory with the $k$ single-trace operators from which it is constructed.  This means that these OPE coefficients must show up at a lower order in $1/N$ in a  $(2+k)$-pt CFT correlator of single-trace primaries.  These statements are analogous to the fact that branch cuts in perturbative scattering amplitudes are never the lowest-order manifestation of interactions.  If we work to leading order in perturbation theory, we can neglect all poles in the Mellin amplitude from multi-trace operators.

Therefore to leading order in $1/N$, we need only consider poles in the Mellin amplitude due to the exchange of the single-trace primary operators.  Crucially, these poles all come from a finite number of conformal blocks, since the spectrum contains only a finite list of single-trace primaries, according to our criterion {\bf 2}.  In the case of external scalar primaries, the conformal blocks take the form of equation (\ref{eq:ConformalBlock}); most other blocks can be constructed via differentiation, as shown in \cite{Costa:2011dw, Costa:2011mg}.  We will discuss higher-spin external operators further in section \ref{sec:HigherSpin}, but for the remainder of this section we will focus on external scalars for simplicity.  The residues of the poles are fixed by unitarity to be proportional to products of the OPE coefficients  $C_{ijk}^a$ from equation (\ref{eq:3ptOPECoefs}).  For example, a spin $\ell_5$ single-trace operator $\CO_5$ exchanged between scalar operators $\CO_1$, $\CO_2$ and $\CO_3$, $\CO_4$ will contribute 
\be
M_{1234}(\delta_{ij}) \supset C_{125} C_{345} B_{\Delta_5}^{\ell_5} (\delta_{ij}).
\ee
If the dimensions of these operators are generic, this will produce an infinite tower of poles in the Mellin amplitude for the $1234$ correlator, corresponding to the exchange of the primary $\CO_5^\ell$ and all its descendants.  But the positions and residues of all of these poles are completely fixed in terms of $C_{125}$ and $C_{345}$.

A finite sum of conformal blocks cannot satisfy our criteria {\bf 3}.  However, as we explained in section \ref{sec:ConformalBlocksandDiagrams} and showed in the case of spin-$0$ and spin-$2$ conformal blocks, there is an elegant connection between conformal blocks and AdS Feynman diagrams.  If we simply express the blocks as a sum over poles, and then drop all other pieces, we obtain convergent expressions that are polynomially bounded.  These expressions are the Mellin amplitudes associated with AdS Feynman diagrams involving the exchange of spin $\ell$ particles dual to CFT operators of dimension $\Delta$.

We have accounted for all possible poles in the 4-pt Mellin amplitudes at leading order in $1/N$, and seen that they must be associated with a finite number of AdS exchange diagrams.  Given our criteria {\bf 3}, this fixes the 4-pt Mellin amplitudes $M_{ijkl}$ up to polynomials.  We saw in section \ref{sec:MellinReview} that at least in the case of scalars, polynomial Mellin amplitudes are in one-to-one correspondence with derivative operators in an AdS Lagrangian; higher spin examples have been constructed in \cite{Paulos:2011ie} and using the methods of \cite{Costa:2011dw, Costa:2011mg}.  Thus we have shown that the criteria we have imposed on the CFT allow us to construct an AdS field theory that reproduces the CFT correlation functions to leading order in $1/N$.  The AdS Lagrangian will involve fields dual to the single-trace operators in the CFT and 3-pt and 4-pt interaction terms that give rise to tree-level exchanges and contact interactions that fully reproduce the perturbative CFT correlators.  

One could of course study higher-point correlators, conformal blocks, and Mellin amplitudes in order to obtain an AdS Lagrangian with general $n$-pt interactions, but this technical extension will be beyond the scope of our analysis.  We will merely point out that by using the conformal block decomposition repeatedly, all conformal blocks for higher $n$-point functions can be reduced to conformal blocks for 4-point functions (often of multi-trace operators).  Equivalently, one can isolate the contribution from operators being exchanged in a single channel by using conglomeration \cite{Unitarity} to turn any $n$-point function of single-trace operators into a 4-point function of multi-trace operators;  explicit knowledge of the four-point function conformal blocks can then be applied in order to extract information about the bulk EFT.  In the next section we will argue that our results extend to all orders in $1/N$.

\subsection{Extension to All Orders in $1/N$}
\label{sec:HigherOrders}

\begin{figure}
\begin{center}
\includegraphics[width=0.85\textwidth]{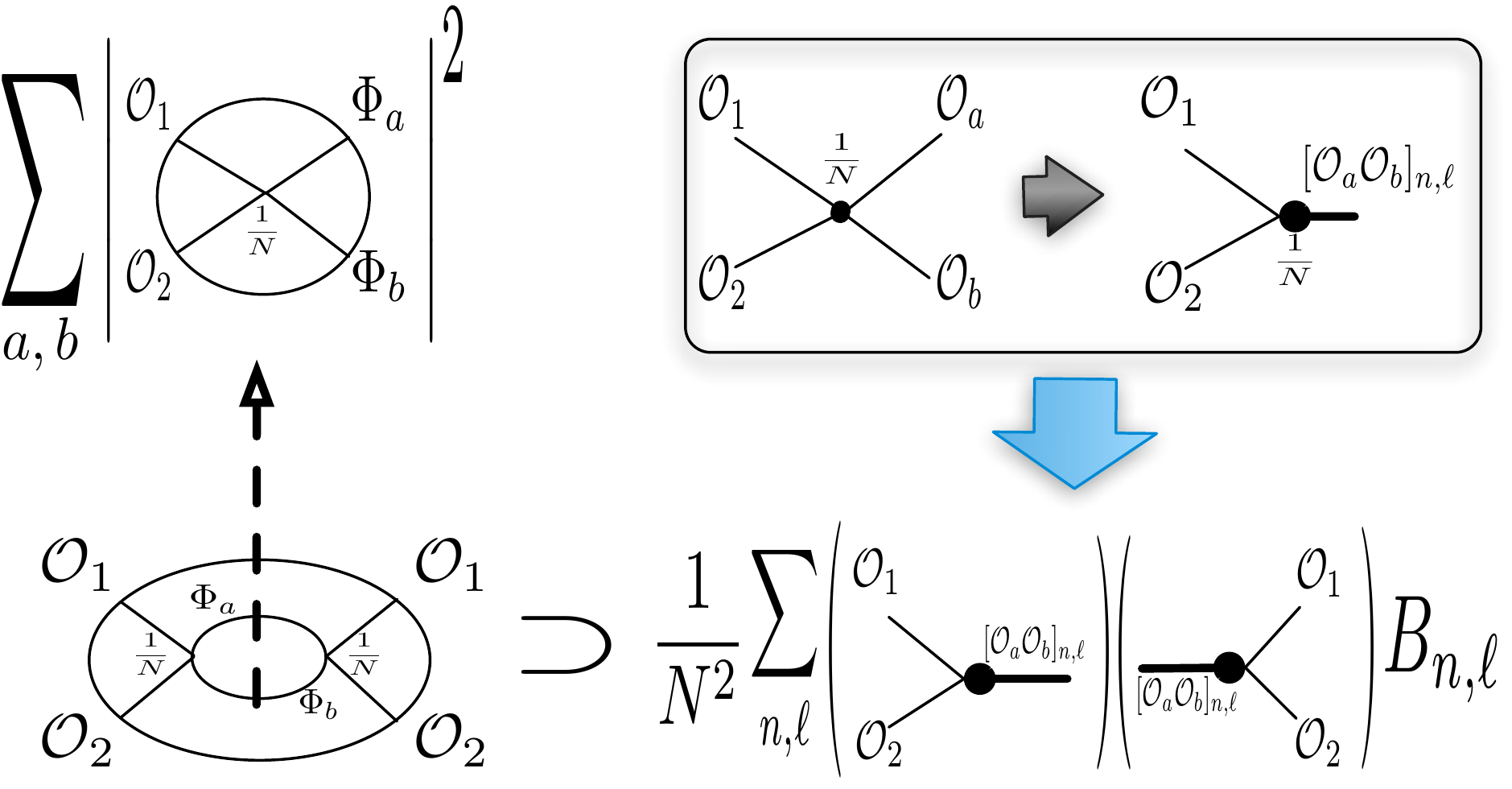}
\caption{This figure gives a schematic depiction of the relationship between the AdS cutting rules (left) and the unitarity or bootstrap equations (right) in the CFT \cite{Unitarity}.  Both connect one order in $1/N$ perturbation theory to the next, so once the AdS and CFT theories agree to leading non-trivial order in $1/N$, their difference will be strongly constrained to all orders.  }
\label{fig:UnitarityRelations}
\end{center}
\end{figure}

We will now explain how our results extend to all orders in $1/N$, first giving a direct constructive proof, and then demonstrating how it relates to quantum mechanical unitarity relations in AdS/CFT \cite{Unitarity}. 

Our constructive methods extends to higher orders in $1/N$ because the meromorphy and polynomial boundedness of the Mellin amplitude preclude any complicated behavior, such as branch cuts, that could not be reproduced through particle exchange and local interactions.  It is this highly constraining feature of Mellin amplitudes (not shared by scattering amplitudes in flat space) that allows one to constructively build an AdS EFT at arbitrary order in $1/N$.  We will sketch the procedure at sub-leading order for four-point functions; the generalization to higher orders is straightforward.  

Assume that we have already applied the construction from the previous subsection to obtain a tree-level AdS effective Lagrangian that reproduces the CFT correlators at leading order in $1/N$.  Now, we turn to the four-point function in the CFT at sub-leading order, say $1/N^2$.  In the simplest possible case, no new AdS interactions will be needed.  For instance, if the bulk theory happens to be $\frac{\lambda}{N} \phi^4$, then the sub-leading four-point function will be exactly the result of standard 1-loop diagrams in AdS from a double insertion of the interaction.  However, in general things may not be so trivial; once we calculate the 1-loop contributions, we may find that there are additional pieces in the Mellin amplitude:
\be
\CA_4^{(\rm sub-leading)} &=& \CA_4^{(\rm 1-loop)} + \CA_4^{(\rm extra)}.
\ee
Such an $\CA_4^{(\rm extra)} $ instructs us to include additional local terms in our AdS effective action with coefficients suppressed by the appropriate powers of $1/N$.  For instance, we may find that $\CA_4^{(\rm extra)}$ contains poles, in which case we simply add cubic interactions proportional to $1/N$ to a field $\phi_2$ with the necessary mass.    After accounting for such poles, by assumption the remaining Mellin amplitude can only be a polynomial, which is then reproduced by shifting the coefficients of contact interactions at order $1/N^2$.  What {\em cannot} happen is that we find that an infinite number of additional fields are required to reproduce $\CA_4^{(\rm extra)}$, since this would violate our assumption {\bf 2} that there are only a finite number of single-trace primary states.  This is a non-trivial statement about the structure of the Mellin amplitude $\CA_4$, because $\CA_4^{(\rm 1-loop)}$ contains an infinite tower of poles in the Mellin amplitude itself\footnote{That is, these poles are not just in the Mellin integrand, as they were at leading order.} coming from double-trace conformal blocks.  Thus, in order to have an AdS EFT dual at subleading order in $1/N$, all but a finite subset of the infinite tower of conformal block poles in $\CA_4^{(\rm subleading)}$ must exactly match $\CA_4^{(\rm 1-loop)}$.  As we explained above, this is enforced by our conditions {\bf 2} and {\bf 3} and the meromorphy of the Mellin amplitude.

Unitarity affords us a conceptual understanding of how our leading order argument extends to higher orders.  
 The point is schematically depicted in figure \ref{fig:UnitarityRelations}.  Once the AdS and CFT theories are fixed at order $1/N$, unitarity in both theories independently forces them to include identical contributions at the next order.  In the AdS effective field theory, unitarity manifests itself through the cutting rules, suitably modified for AdS, which connect the imaginary parts of $L$-loop diagrams to phase space integrals (or sums) over products of lower order diagrams.  In the CFT, unitarity manifests itself through the famous bootstrap relations, which use the OPE to connect the conformal block decomposition at one order in perturbation theory to the same decomposition at the next order.  We showed how this works in the flat space limit of AdS/CFT in \cite{Unitarity}, obtaining the optical theorem for the S-Matrix, but the same relations also hold for any finite AdS scale $R$.  Of course unitarity may not fully determine higher order terms in perturbation theory, because there may be new, more weakly coupled particles and new UV sensitive counter-terms that must be fixed, as we discussed in the few paragraphs.  However, unitarity provides a powerful supporting argument that by including minor sub-leading corrections, we can obtain an AdS EFT dual to any CFT satisfying our three criteria at all orders in $1/N$.

\section{Perturbative Unitarity and the Necessity \\ of Polynomial Boundedness}
\label{sec:Unitarity}

So far we have taken the polynomial boundedness criterion {\bf 3} as an assumption and discussed the implications.  However, we would clearly like to understand which CFTs actually satisfy this criterion, and why.  Are there more fundamental principles from which this criterion can be derived? 
In this section, we will give evidence that suggests (though does not prove) that for CFTs with a $1/N$ expansion and a large gap $\Delta_\Lambda$ in scaling dimensions\footnote{Here by a `gap' we mean that we are neglecting operators of dimension larger than $\Delta_\Lambda$, but we are certainly not assuming that the Mellin amplitude has any sort of expansion in $1/\Delta_\Lambda$ -- in fact this is the statement we would like to prove!} of single-trace operators, the requirement of unitarity may enforce polynomial boundedness.  

Intuitively, the connection between unitarity and boundedness is familiar from constraints on amplitudes in flat space.  The optical theorem combined with limits on the growth of cross sections, such as the Froissart bound, imposes asymptotic polynomial boundedness on scattering amplitudes in the physical region, with a definite polynomial degree.  Passing to an effective field theory description, these exact unitarity bounds turn into perturbative unitarity limits that predict the breakdown of perturbation theory at a certain scale.  If the EFT is to provide a perturbative and unitary description of the physics at energies below a `gap' scale $\Delta_\Lambda$, then we obtain constraints on the allowed interactions and their strength, with irrelevant operators forced to have an EFT type expansion.  This is analogous to our situation because by assumption we have perturbative and unitary CFT correlators for operators with scaling dimensions below a gap.

\begin{figure}
\begin{center}
\includegraphics[width=0.75\textwidth]{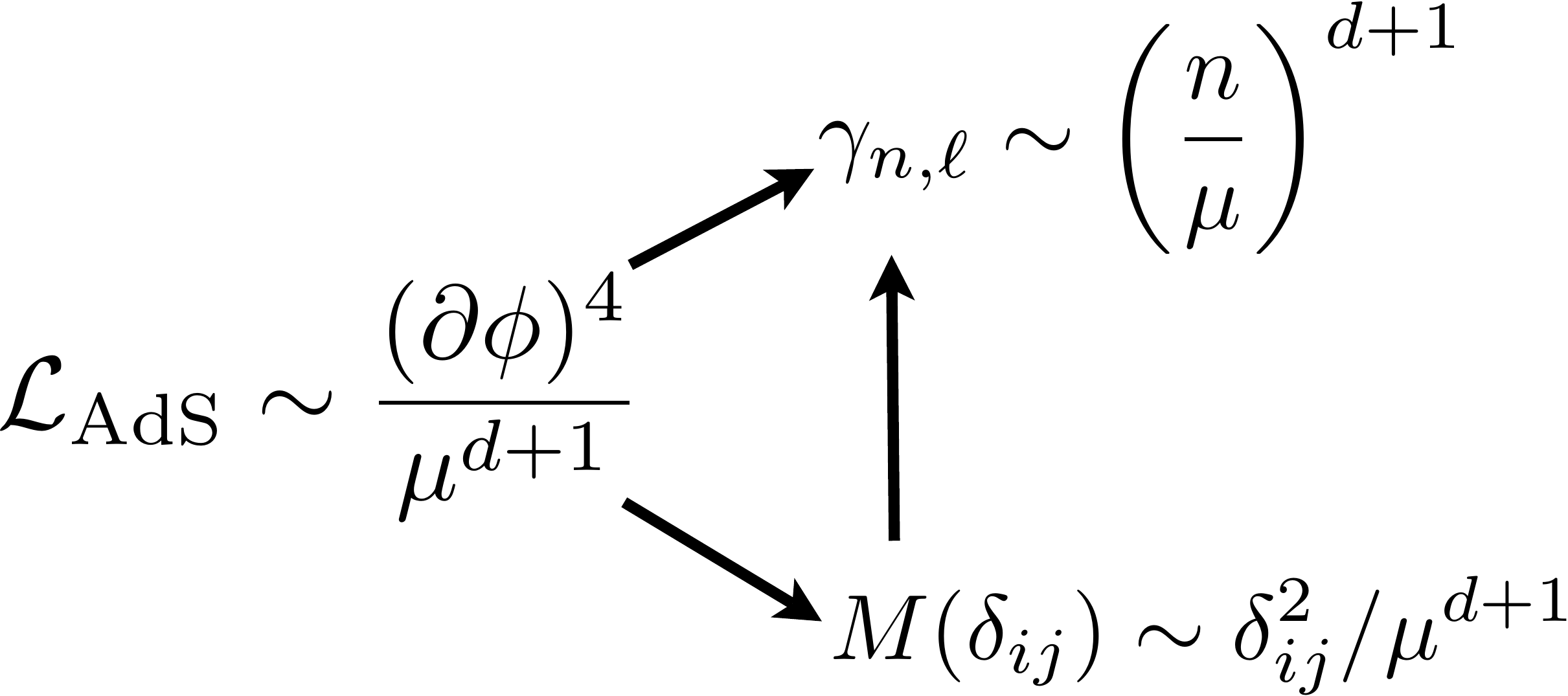}
\caption{Irrelevant interactions in AdS like $(\partial \phi)^4$ create both Mellin amplitudes that grow at large $\delta_{ij}$ and anomalous dimensions of double-trace operators $\CO_{n,\ell}$ that grow at large dimension, eventually violating unitarity in perturbation theory.  Alternatively, starting only with the Mellin amplitude, one can directly calculate the anomalous dimensions without a priori identifying the corresponding bulk interaction. } 
\label{fig:AnomDimsAndMellin}
\end{center}
\end{figure}

In \cite{Katz} a direct perturbative unitarity bound was discovered for CFTs, and it can be used to powerfully constrain the growth of Mellin amplitudes at large $\delta_{ij}$.  The anomalous dimensions $\gamma_{n,\ell} $ of double-trace operators $\CO_{n,\ell}$ are bounded \cite{Katz} by 
\be
|\gamma_{n,\ell}| < 4 \qquad (n\gg 1).
\label{eq:gammabound}
\ee
If an anomalous dimension exceeds this bound then some double-trace CFT operators $\CO_{n,\ell}$ create states with negative norms at leading order in perturbation theory.  The bound also has an interpretation in terms of scattering, because $\gamma_{n,\ell}$ are related to scattering amplitudes in the flat space limit \cite{Katz}.  In order for this violation of unitarity to be resolved perturbatively, new CFT operators with dimension less than $\Delta[\CO_{n,\ell}]$ must be ``integrated in'' to the effective conformal theory and mix with $\CO_{n,\ell}$.   As an illustrative example, consider the effective conformal theory corresponding to $ \phi^4/\mu$ theory in AdS$_5$.  The interaction is irrelevant, and therefore becomes an increasingly large perturbation for operators of large dimension; in this case, the anomalous dimensions of double-trace operators grow linearly as a function of $n$, $\gamma_{n,\ell} \sim n/(\mu R)$, and thus violate the bound in equation (\ref{eq:gammabound}) around $n \approx \mu R$.    In general, the growth of anomalous dimensions $\gamma_{n,\ell}$ at large $n$ can be easily read off from the AdS interaction by dimensional analysis, treating $n$ as an energetic parameter \cite{Katz}.  

We have also seen how local AdS interactions produce Mellin amplitudes whose growth can be read off from counting powers of fields.  This connection is depicted in Figure \ref{fig:AnomDimsAndMellin} for the example $(\partial \phi)^4$.  In fact, knowledge of the specific AdS interaction is not a priori necessary in this case (though it is useful and illuminating), since one can directly calculate the corresponding multi-trace anomalous dimensions; at a technical level, the  most efficient method is that of ``conglomerating'' operators developed in \cite{Unitarity}.  However, there is also a direct and simple line of reasoning.  The $\Gamma(\delta_{ij})$ factors in the Mellin integrand produce a great many poles at negative integral $\delta_{ij}$, and in perturbation theory the single poles can coincide, producing double poles at generic negative integral $\delta_{ij}$.  Double and higher poles are absent from the exact Mellin amplitude, but they are present in perturbation theory, producing the position-space logarithms that signal the presence of anomalous dimensions.  If the Mellin amplitude behaves smoothly in the physical region near these double poles, its normalized value approximates the size of the anomalous dimensions. 

 Thus we expect that large powers of $\delta_{ij}$ in the Mellin amplitude that are not accompanied by appropriate powers of the gap $\Delta_\Lambda$ in the denominator will lead to violations of unitarity at low dimensions $\Delta < \Delta_\Lambda$, contrary to criterion {\bf 1}.  For instance, a Mellin amplitude of the form $M(\delta_{ij}) \sim \exp(-\delta_{12})$ causes violations of unitarity at dimensions of order one, producing a cutoff for the AdS effective theory that is no larger than the curvature scale $1/R$. 

However, this argument does not immediately rule out Mellin amplitudes with rapid growth occurring away from the physical regime (which is e.g. real $\delta_{12} < -\frac{\Delta_1 + \Delta_2}{2}$ in the 4-pt amplitude, viewed in the $s$-channel).  For instance, as we have seen, a single conformal block will not satisfy criterion {\bf 3}, but it is perfectly unitary in the s-channel, and the fact that it violates unitarity becomes apparent only when it is decomposed in the cross-channel, where one finds that no conformal block decomposition is possible.  Therefore one cannot show that ECTs satisfying criteria {\bf 1} and {\bf 2} actually also satisfy {\bf 3} without a more subtle argument that uses crossing-symmetry and/or higher-point correlators to extend the bound on the growth of $M(\delta_{ij})$ to apply outside the physical region.  This is a very interesting subject for future investigation.  

\section{Flat Space Limit and S-Matrix Analyticity}
\label{sec:FSLandAnalyticity}

Recent work has emphasized the profound similarity between Mellin space for CFTs and momentum space for Scattering Amplitudes \cite{JoaoMellin, NaturalLanguage, Paulos:2011ie}.  It has also been shown that in the flat space limit of AdS, the Mellin amplitude basically becomes the S-Matrix \cite{JoaoMellin, NaturalLanguage, Analyticity, Unitarity}, with the Mellin space variables re-interpreted as physical Mandelstam invariants.  Therefore, our arguments have analogs in the much more familiar realm of scattering theory.  We will begin this section by describing how the S-Matrix version of our argument would work; this may be especially useful to readers who are unfamiliar with Mellin space.  

Imagine that we have a perturbative, analytic\footnote{up to poles and branch cuts corresponding to single and multi-particle states.}, unitary, Poincar\' e invariant S-Matrix for a finite number of species of particles, and that the S-Matrix elements are polynomially bounded at large momentum.  More generally, the S-Matrix could simply have an expansion in energies divided by some cutoff scale $\Lambda$.  As long as we study energies $E \ll \Lambda$, we can work to any given accuracy by ignoring all terms beyond a certain order, leaving us with scattering amplitudes that are effectively polynomially bounded.

We assumed a perturbative expansion, so let us call the expansion parameter $\frac{1}{N}$.  Focusing on the 2-to-2 scattering amplitude for some particle species $\phi$, the only non-analyticity will be from poles and branch cuts.  However, the branch cuts are an effect beyond the leading order in $\frac{1}{N}$, because by unitarity they are related to the product of two non-trivial scattering amplitudes.  So at leading order we can only have poles from single-particle states, and again by unitarity these poles must be located at the physical masses $m_i^2$ of particle states in our theory.  Subtracting these pole terms, we are left with an entire polynomially bounded function of the momenta, so of course it must be a polynomial.  We can build a flat space Lagrangian with 3-pt vertices that give rise to the pole terms, and contact interactions that produce the pure polynomial terms in the momenta, so we see that the S-Matrix must have a description as an effective quantum field theory.

A subtle point is our use of analyticity.  Very naively, all that unitarity tells us is that the S-Matrix can have imaginary contributions
\be
M(s,\cos \theta) \supset i \frac{1}{N^2} P_\ell ( \cos \theta ) \delta(s - m^2),
\ee
corresponding to the physical particle states in our theory with spin $\ell$ and mass $m$.   This is directly analogous to the presence of a single-trace conformal block in the CFT correlator; it is literally a single block in the flat space limit of AdS/CFT \cite{Unitarity}.  In the case of the S-Matrix, analyticity suggests that these delta functions should be replaced by poles, but it is not completely obvious that these poles can only be simple propagators and not more complicated functions of the momenta.  However, our CFT argument in section \ref{sec:ConformalBlocksandDiagrams} is much more direct, because conformal invariance and CFT unitarity require a meromorphic Mellin amplitude with only simple poles.  In other words, because analyticity is a somewhat murky stand-in for locality, our derivation of AdS field theory from the CFT is actually conceptually cleaner than the analogous derivation of a flat space EFT from properties of the perturbative S-Matrix.

Now let us move on and discuss how S-Matrix analyticity arises from AdS field theory, and why it breaks down when our criteria are violated.  Confirming a conjecture of Penedones \cite{JoaoMellin}, we showed \cite{Analyticity} that the scattering matrix $T = S - 1$ can be obtained from the Mellin amplitude via the contour integral and limit
\be
\label{eq:FSLFormula}
T(s_{ij}) =  \lim_{R \to \infty} \frac{1}{\CN} \delta^{d+1} \left( \sum_i p_i \right) \int_{-i\infty}^{i\infty} d \alpha \ \!  e^{\alpha}  \alpha^{h - \sum_i \frac{\Delta_i}{2}}  
M \left( \delta_{ij} = -\frac{R^2 s_{ij}}{4 \alpha} \right),
\ee
where the contour in $\alpha$ is taken to the right of all the poles in the integrand, and $\CN$ is a normalization factor \cite{JoaoMellin, Analyticity}.  We see that the Mellin space variables are set proportional to the physical Mandelstam invariants of the scattering process, and that it is primarily information about the large $\delta_{ij}$ behavior of the Mellin amplitude that determines the S-Matrix.  

The reader can check via a very straightforward calculation \cite{JoaoMellin} that any polynomial in the Mellin variables will produce an identical polynomial in the Mandelstam invariants as long as the coefficients (couplings) scale appropriately in the flat space limit.  It has been shown before in several different ways \cite{JoaoMellin, NaturalLanguage, Analyticity} that Mellin-space propagators turn into momentum space propagators in the flat space limit of AdS.  Furthermore, in \cite{JoaoMellin, Analyticity} various one and two loop computations have been performed in AdS, and one can see how a coalescence of Mellin space poles produce the familiar branch cuts of the loop-level S-Matrix.  A slightly more involved calculation shows how the exchange of unstable particles in AdS leads to poles in the complex $s$-plane in the flat space limit.  Thus equation (\ref{eq:FSLFormula}) reproduces the familiar analyticity properties of the  S-Matrix in the flat space limit of AdS field theory.

What about CFT correlators that do not have a nice interpretation in terms of AdS field theory?  A natural example are the conformal blocks themselves, which grow exponentially at large $\delta_{ij}$.  We showed in \cite{Analyticity, Unitarity} that the flat space limit of a conformal block is
\be
B_{\Delta, \ell} \to P_\ell(\cos \theta) \delta (s - m^2),
\ee
up to a normalization factor.  Clearly this is not an analytic function of the momenta, as expected.  We already saw how dropping the exponentially growing parts of a conformal block results in a AdS exchange diagram, but perceptive readers might wonder what would happen if instead we were to preserve exponential boundedness by jettisoning unitarity.  For this purpose we can study the polynomially bounded Mellin-space function
\be
S_{\Delta, \ell} = \frac{\Gamma \left(\frac{\Delta-\ell-\delta}{2} \right) \Gamma \left(\frac{2h-\Delta - \ell - \delta}{2} \right) }{ \Gamma \left(\Delta_a-\frac{\delta}{2} \right) \
 \Gamma \left(\Delta_b - \frac{\delta}{2} \right) } P_{\ell}(\delta_{ij}) .
\ee  
This function is a linear combination of a conformal block and its non-unitary `shadow' with dimension $d - \Delta$;\footnote{If $h-1 < \Delta < h$, then $d-\Delta$ is above the unitarity bound, but the shadow field contribution is still non-unitary because its conformal block contributes to $S_{\Delta,\ell}$ above with a negative sign.}  more simply it can be obtained from our equation (\ref{eq:ConformalBlock}) by dropping the exponential prefactors.  As long as $\Delta_a + \Delta_b  >  h$ the function $S_{\Delta, \ell}$ vanishes as $\delta \to \infty$ and depends on the other Mellin variables through the polynomial $P_\ell(\delta_{ij})$, so we can write $S_{\Delta, \ell}$ as a convergent sum over its poles.  This form expresses $S_{\Delta, \ell}$ as the difference between two AdS propagators.  In the flat space limit, these two propagators have squared masses that differ only by $\CO(1/R)$, the AdS scale.  The analytic pieces of these propagators cancel in the difference, and we are left with unitarity-violating imaginary delta functions in the S-Matrix.  Thus cancellations between different conformal blocks can result in minor violations of analyticity, although this is highly non-generic and probably impossible in unitary theories.

One might also wonder what happens with other CFTs that violate our criteria.  An obvious example are the minimal models, which clearly violate our first two criteria because they are not perturbative and, when the Virasoro primaries are decomposed in primaries under the global conformal group, have a large density of states.  Interestingly, however, an exact Mellin amplitude for the 2-d Ising model has been derived \cite{BaltJoao}, and it does obey our third criterion of polynomial boundedness.  Of course it does not make any sense to take the flat space limit of AdS for this Mellin amplitude, but formally proceeding with equation (\ref{eq:FSLFormula}) gives a vanishing result.   In fact it may be the case that many CFTs have polynomially bounded exact Mellin amplitudes for their correlators, but this does not necessarily mean that these Mellin amplitudes can be well-approximated by effective CFTs with a parametrically large gap in dimensions.  

A fascinating final issue involves the relationship between our boundedness criterion, the usual exponential boundedness assumptions about the analytic S-Matrix, and flat space scattering amplitudes with black holes as intermediate states.  As we discussed in \cite{Analyticity}, a well-known issue is that the 2-to-2 S-Matrix at trans-Planckian energies and small impact parameter should be suppressed by $e^{-S_{BH}}$, where $S_{BH}$ is the entropy of a black hole of mass equal to the center of mass energy.  But this function violates exponential boundedness, because for small black holes $S_{BH} \sim E^r$ with $r > 1$.  However, for any finite AdS scale $R$, at very large energies one eventually moves into a regime where large AdS black holes will be produced, in which case one returns to a regime where $r \leq 1$, with exponentially bounded scattering amplitudes.  Thus it may be that when the flat space S-Matrix is viewed as a holographic object derived from AdS/CFT, it only truly violates asymptotic exponential boundedness in the strict $R \to \infty$ limit, but is well-behaved at all finite $R$. Violations of exponential boundedness may ultimately be viewed as transients associated with parametric limits (large $N$, $\lambda$, etc) of an otherwise much better behaved Mellin amplitude.

\section{Constraints on Higher-Spin Operators}
\label{sec:HigherSpin}

One might have naively expected large central charge CFTs to have low dimension `single-trace' operators of all spins.  After all, such operators certainly exist in weakly coupled gauge theories, where one can construct single-trace operators with arbitrary numbers of tensor indices from gauge fields and derivatives.  From this point of view, it is surprising that CFTs with AdS field theory descriptions can avoid having low-dimension single-trace operators with spin $\ell > 2$.  

In fact, bulk field theories of higher spin particles must endure various well-known constraints and pathologies (for a review and references see e.g. \cite{Weinberg:1995mt, Porrati:2008rm}) that help to explain their absence from low-energy physics.  In this section we will explain that one of the most famous constraints on higher spin interactions, the Weinberg soft theorems \cite{Weinberg:1965nx}, can be understood within the boundary CFT.  This goes a long way towards providing an intrinsic CFT explanation why our criteria automatically tend to restrict the spin of single-trace operators.

Let us begin by reviewing Weinberg's theorems, which are based on studying soft emission in scattering processes and imposing Lorentz invariance on the S-Matrix.  Consider an $n$-particle scattering process that emits a soft massless boson of spin $\ell$ and momentum $q$.  In the limit that $q \cdot p_i \ll p_i \cdot p_j$, where $p_i$ are the momenta of the $n$ other particles, the full scattering amplitude can be approximated by
\be
\label{eq:SoftLimitScattering}
\mathcal{M}_{n+1} \approx \mathcal{M}_n \sum_{i=1}^n g_i \frac{\epsilon_{\mu_1 ... \mu_\ell} p_i^{\mu_1} \cdots p_i^{\mu_\ell} }{q \cdot p_i},
\ee
where $\epsilon_{\mu_1 ... \mu_\ell}$ is the transverse, traceless polarization tensor of the soft particle.  This formula corresponds to a sum of Feynman diagrams where the soft particle is radiated from the external legs of the diagram, producing a large propagator factor in the soft limit.  The divergence as $q \to 0$ corresponds in position space to a diverging integral over the time of interaction between the $i^{\mathrm th}$ particle and the soft boson.  This suggests that this soft factorization result follows directly from considerations of locality, although we will not attempt to make this precise.  The $n$ hard particles can have arbitrary spin, so some of the factors of momenta $p_i$ in the numerator can be replaced with operators constructed from the spins of the hard particles, but this complication does not change the final result.

This simple form for the S-Matrix leads to powerful constraints due to the Lorentz transformation properties of the polarization tensor.  Under Lorentz transformations, massless polarization `tensors' do not transform as pure tensors; instead they also shift by quantities such as $q_{\mu_1} v_{\mu_2... \mu_\ell}$.  In other words, they transform like generalized gauge fields under gauge transformations, and so to preserve Lorentz invariance we must have
\be
\label{eq:SoftThmConstraint}
0 = \sum_{i=1}^n g_i \ \! p_i^{\mu_2} \cdots p_i^{\mu_\ell} .
\ee
This is a highly constraining relation, and the fact that it can be solved at all is non-trivial.  When $\ell = 1$ it can be satisfied only if the sum of the 3-pt couplings $g_i$ vanishes, so massless spin one particles must couple to a conserved current.  In the case $\ell = 2$ the sum can vanish for all $p_i$ only if all of the $g_i$ are identical; this implies the universality of low-energy gravitational interactions.  However, when $\ell > 2$ the condition imposes a kinematic constraint on the S-Matrix that is independent of momentum conservation, so massless higher spin particles cannot produce couplings that lead to long-range forces.

The Weinberg soft theorem essentially follows from considerations of Lorentz invariance, unitarity, and locality as applied to the S-Matrix, which is the holographic observable in flat spacetime.  When we compute the S-Matrix using a bulk field theory in flat space, the soft theorems constrain that field theory.  We will now argue that a generalization of Weinberg's argument generically applies to CFT correlators satisfying our three criteria.  This will imply constraints on AdS effective field theories dual to such CFTs, but the constraints can also be viewed as purely holographic, applying directly to the CFT itself in the same way that Weinberg's theorem applies directly to the S-Matrix.

The idea behind generalizing the soft theorems is very simple.  CFTs with AdS EFT descriptions with a cutoff $\Lambda$ parametrically larger than the AdS scale $1/R$ include physical processes dual to short-distance physics in the bulk.  But at very short distances, the difference between AdS and flat space becomes negligible, and so the usual Weinberg soft theorems must obtain.  This familiar line of reasoning also justifies recent results on the holographic S-Matrix \cite{Analyticity}.  The question that remains is how to justify the soft theorems with finite $\Lambda R$ using practical AdS/CFT technology for expressing CFT correlation functions.

\begin{figure}
\begin{center}
\includegraphics[width=\textwidth]{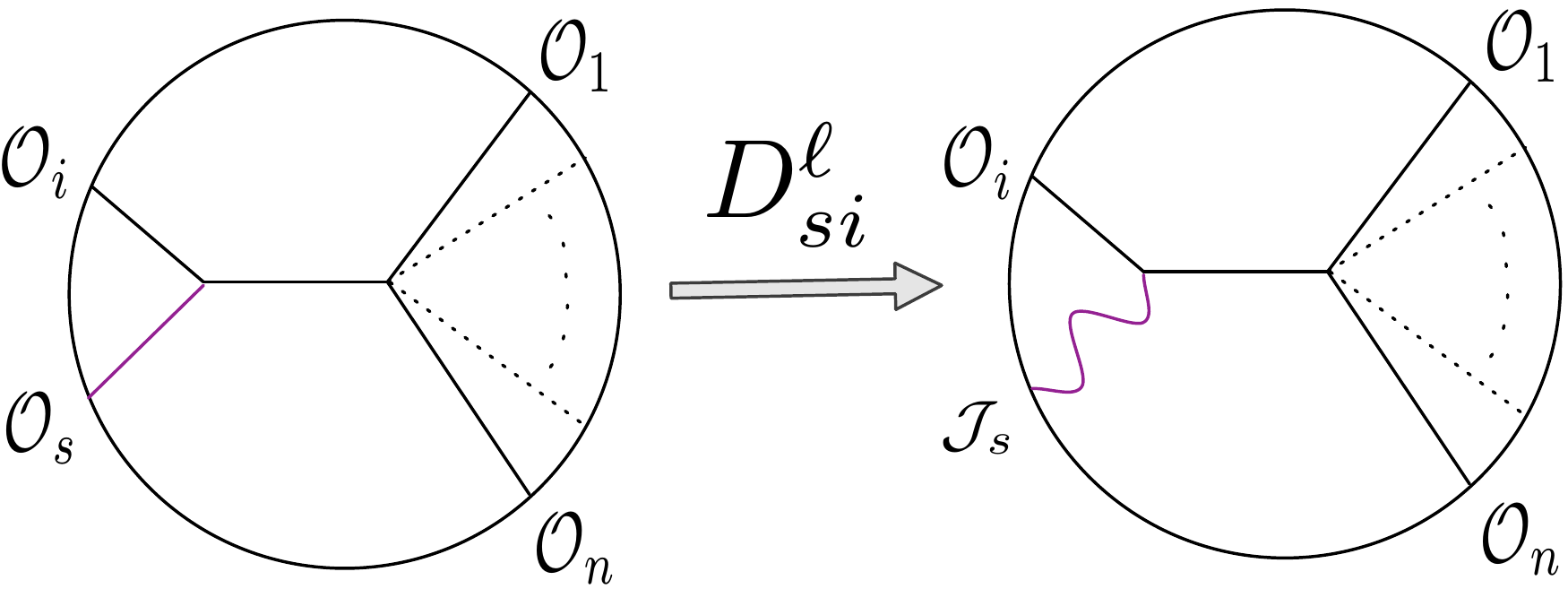}
\caption{This figure shows how the differential operator $(D_{si})^\ell$ maps a certain class of scalar CFT correlation functions into correlators with a current $\CJ_s$.  These are also the AdS Feynman diagrams relevant for the AdS/CFT analog of the Weinberg soft limits.  Imposing current conservation on the sum of these diagrams provides strong constraints on the couplings of conserved higher spin currents. }
\label{fig:SoftLimitDiagrams}
\end{center}
\end{figure}

\subsection{Warm-up: Spin-1}
\label{sec:Spin1Soft}

Let us begin by generalizing equation (\ref{eq:SoftLimitScattering}) in the case of spin-1 emission to AdS/CFT.  We will study a bulk process involving an n-point scalar interaction and dress it by emitting a soft gauge boson.  This produces an $(n+1)$-point CFT correlation function of a current $J_\mu$ with $n$ charged scalar operators.

How can we compute this CFT correlator?  There are at least two options:  we could use the explicit Feynman diagram techniques from \cite{Paulos:2011ie}, or  we could differentiate a 5-pt amplitude of scalars using the technology of \cite{Costa2011mg, Costa:2011dw}.  We will use the latter technique, because it naturally accords with our observations above concerning the relationship between conformal blocks and AdS Feynman diagrams.  

The idea of \cite{Costa:2011dw} is that 3-pt functions of tensor fields can be constructed from a basis formed by differentiating scalar 3-pt functions.  Since conformal blocks are basically defined by the OPE limits of their two pairs of external operators, this means that we can construct most conformal blocks involving external spinning operators by taking linear combinations of derivatives of scalar conformal blocks.  We saw in section \ref{sec:ConformalBlocksandDiagrams} that conformal blocks can be used to immediately construct AdS exchange diagrams, so in general we can use the methods of \cite{Costa:2011dw} to compute AdS exchange diagrams including external operators with spin.  This is how we will compute the diagrams in figure \ref{fig:SoftLimitDiagrams}.

The derivative operator we will need for this example can be written as \cite{Costa:2011dw}
\be
\label{eq:SpinRaising}
D_{si} = (P_s \cdot P_i) (Z_s \cdot \nabla_i)  - (Z_s \cdot P_i) (P_s \cdot \nabla_i)  \ \ \ \mathrm{with} \ \ \ \nabla_i = \frac{\partial}{\partial P_i}.
\ee
This operator, when applied to a scalar correlator dual to an AdS Feynman diagram where operators $\CO_s$ and $\CO_i$ connect to the rest of the correlator via a scalar exchange, turns that correlator into that of a current $J_s$ with the other operators, as in figure \ref{fig:CurrentConservation}.   The full correlator can then be written as
\be
Z^A \left\langle \CJ_A (P_s)  \CO (P_1) \dots \CO^\dag (P_n) \right\rangle = \sum_{i=1}^n q_i  D_{si} W(si \to 1 \cdots \hat i \cdots n) ,
\label{eq:CurrentScalarCorrelator}
\ee
where $q_i$ is the $i^{\rm th}$ operator charge and $W(si \to 1 \cdots \hat i \cdots n)$ is the scalar Feynman diagram depicted on the left in figure \ref{fig:SoftLimitDiagrams}, with propagation of fields $s$ and $i$ to the other $n-1$.  The scalar $\CO_s$ must have dimension $\Delta_s = d$ in order for the action of $D_{si}$ to obtain a conserved current $\CJ_s$ with dimension $d - 1$.

 These Feynman diagrams can be trivially computed using the AdS Feynman rules \cite{NaturalLanguage, Paulos:2011ie, Nandan:2011wc, Analyticity}, and take the form
\be
W(si \to 1 \cdots \hat i \cdots 4) = \sum_{m=0}^\infty \frac{R_m}{\Delta_s + \Delta_i - 2 \delta_{si} - (\Delta_i + m)}
\ee
in Mellin space, where the residues $R_m$ are independent of all the $\delta_{ab}$.  Applying the derivative operator $D_{si}$ to the Mellin-space kinematic invariant $\prod_{a<b} P_{ab}^{-\delta_{ab}}$ is straightforward, and so we obtain the  Mellin representation for the correlator
\be
\label{eq:CurrentScalarCorrelator}
& Z^A & \! \! \!  \left\langle \CJ_A (P_s)  \CO (P_1) \CO (P_2) \CO^\dag (P_3) \CO^\dag (P_4) \right\rangle  \\ 
&& \ \ =  \sum_{i \neq j} \left[\frac{Z_s \cdot P_j}{P_s \cdot P_j} -  \frac{Z_s \cdot P_i}{P_s \cdot P_i}\right] \delta_{si} \delta_{sj}  W_{ij}^{si, sj}(si \to 1 \cdots \hat i \cdots 4) ,\nn 
\ee
where the subscripts and superscripts on the Mellin amplitude $W$ indicate that it is a function of $\delta_{ij} \to \delta_{ij} -1$ and $\delta_{si} \to \delta_{si} +1,\delta_{sj} \to \delta_{sj} +1$, respectively.  
 Notice the similarity in structure to the S-Matrix formula (\ref{eq:SoftLimitScattering}); we have a `hard' scalar amplitude multiplied by a sum of `soft' propagators and a polarization structures.

Weinberg's soft theorems for the S-Matrix follow from Lorentz invariance, and more specifically from the non-trivial Lorentz transformation properties of massless polarization tensors.  Conserved CFT currents are dual to massless AdS gauge fields, so to obtain the CFT equivalent of equation (\ref{eq:SoftThmConstraint}), we need to impose current conservation on the current correlator of equation (\ref{eq:CurrentScalarCorrelator}).  In order to consider current conservation, we will need to know how the  gradient operator $\partial/\partial x^{\mu_1}$ acts on a correlator $\CA^{\mu_1 \cdots \mu_\ell}$ of a symmetric traceless spin-$\ell$ field when the correlator is written using the embedding space formalism with the $Z^A$ polarization vectors.  This action was worked out in \cite{Costa:2011dw} and takes the form of the following  differential operator:
\be 
\label{eq:CurrentConsOp}
\mathcal{D}_s = \nabla^A_s \left[\left(h -1 + Z_s \cdot \frac{\partial}{\partial Z_s} \right) \frac{\partial}{\partial Z^A_s} - \frac{1}{2} Z_{sA}  \frac{\partial^2}{\partial Z _s\cdot \partial Z_s}\right].
\ee
For $\ell=1$, only the term with a single $Z$ derivative contributes.  $\mathcal{D}_s$ can be very simply applied in our example, using the following trick.
Note that
\be
2 \frac{\mathcal{D}_s \cdot D_{si}}{h-1} &=& 2  (P_s \cdot P_i) \left( \delta_{AB} - \frac{P_{sA} P_{iB} }{P_s \cdot P_i }\right) \nabla_i^A \nabla_s^B \\ 
&=& \frac{1}{2} \left( \mathcal{L}_s + \mathcal{L}_i  \right)^2 - P_s \cdot \nabla_s (P_s \cdot \nabla_s - d) - P_i \cdot \nabla_i (P_i \cdot \nabla_i - d) \\
&=& \left[ \frac{1}{2} \left( \mathcal{L}_s + \mathcal{L}_i  \right)^2  - \Delta_i (d - \Delta_i) \right] - \Delta_s(d - \Delta_s)
\label{eq:ConservationDotDerivative},
\ee
where $\mathcal{L}$ are the conformal generators, so the first term in the second and third line is the quadratic Casimir of the conformal group.  The last equality follows because $P_a \cdot \nabla_a$ acts as a dilatation on the operator $\CO_a$ in the correlator.  

The expression in square brackets on the final line in equation (\ref{eq:ConservationDotDerivative}) would annihilate a conformal block of dimension $\Delta_i$ exchanged between $\CJ_s \times \CO_i$ and the remainder of the diagram; when acting on the AdS Feynman diagram it precisely cancels the Mellin-space propagator \cite{NaturalLanguage}.  This follows because the conformal Casimir is also the AdS Laplacian \cite{NaturalLanguage}, and the AdS propagator solves the AdS Laplace equation with a delta function source.  The last term in equation (\ref{eq:ConservationDotDerivative}) vanishes when $\Delta_s = d$, corresponding to the case when the CFT current $\CJ_A$ is conserved.  So we have proven that
\be
\mathcal{D}_s \sum_{i=1}^n q_i  D_{si} W(si \to 1 \cdots \hat i \cdots n) = W(12\dots n) \sum_{i=1}^n q_i ,
\ee
where $W(12 \dots n)$ is the `hard' correlator of the four scalars.   
The right hand side can vanish only when the sum of the charges is zero.  As desired, this is identical to the constraint in equation (\ref{eq:SoftThmConstraint}) for the case that the spin $\ell = 1$.

\subsection{What is the Soft Limit, and when does it break down?}

In the previous section we saw how spin-one soft factors for the flat space S-Matrix can be generalized to CFT correlation functions.  However, thus far we have glossed over a crucial issue -- how do we know that a certain subset of terms in the CFT correlator, corresponding to the specific Feynman diagrams pictured in figure \ref{fig:SoftLimitDiagrams}, dominate the CFT correlation function?  In other words, what is the `soft limit' for AdS/CFT?

The answer, of course, is that the generalization of Weinberg's soft limit to AdS/CFT must be the region of Mellin space where
\be
\delta_{si} \ll \delta_{ij} ,
\ee
where $s$ labels the `soft' particle of spin $\ell$.  As we have repeatedly emphasized \cite{NaturalLanguage, Analyticity}, the Mellin space variables can be interpreted as AdS/CFT versions of the Mandelstam invariants, with identical intuition about the UV, the IR, and factorization.  In the case of the S-Matrix, the subset of Feynman diagrams where the soft particle radiates from an external leg dominate in the soft limit.  Since flat space momenta can be taken to be arbitrarily soft, these diagrams can be made arbitrarily large compared to the rest.  In AdS/CFT the situation is more subtle, although we will see that the end result basically corresponds to studying the S-Matrix soft theorems in the presence of an IR cutoff.  

Mellin space propagators have poles, so naively one might try to put the soft variables $\delta_{si}$ `on-shell', i.e. at the location of a pole.  However, as we discussed in section \ref{sec:MellinPoles}, this would just give the contribution from the exchange of the specific operators associated with that pole.  In other words, the Mellin poles are analogous to the imaginary delta functions from the $i \epsilon$ prescription in flat space propagators\footnote{In fact we showed in \cite{Analyticity, Unitarity} that this is more than an analogy -- conformal blocks become delta functions in the center of mass energy when we take the flat space limit of AdS/CFT.}.   Focusing solely on $\delta_{si}$ at these poles is therefore too restrictive.  

Instead of localizing on poles in the Mellin amplitude, we want to study the entire region in Mellin space where diagrams with poles in $\delta_{si}$ dominate over diagrams without these poles.  The magnitude of the Mellin amplitude (and the diagrams that contribute to it) has a physical meaning because the contour integral over the Mellin variables localizes on poles from the $\prod_{a<b} \Gamma(\delta_{ab})$ factor in the Mellin integrand.  This will produce a large number of different contributions to the CFT correlator with $\delta_{ab}$ localized at various negative integers $n_{ab}$.  We will obtain a large contribution to the correlator when the Mellin amplitude itself is large over a region where the $\delta_{ab}$ vary over a significant range, because $M(n_{ab})$ will be the residue of the Mellin integral.   Therefore, we want to argue that as long as  $\delta_{ij} \gg \delta_{si} \gtrsim 1$, the diagrams with `radiation' from external legs pictured in figure \ref{fig:SoftLimitDiagrams} give by far the largest contribution to the Mellin amplitude.  

For this purpose, we need to keep in mind that the AdS field theory may have a UV cutoff (or equivalently, we are dealing with an Effective CFT \cite{Katz}). 
Given such an AdS EFT cutoff $\Delta_\Lambda R^{-1}$ corresponding to a gap $\Delta_\Lambda$ in the dimensions of CFT operators, we necessarily have $\delta_{ij} < \Delta_\Lambda$.  The soft variables $\delta_{si}$ range over a region of at least $O(1)$ size, so this means that the ratio of hard to soft Mellin variables will be at most of order $\Delta_\Lambda$.  This puts a fundamental restriction on the power of the AdS/CFT generalization of the soft theorems that is equivalent to imposing an IR cutoff of $1/R$ on the flat space S-Matrix.  As usual, this accords with the intuition that AdS/CFT puts quantum gravity in a box.

Assuming a parametrically large cutoff $\Delta_\Lambda \gg 1$, we can justify the dominance of the soft limit, but only after making an assumption about the genericity of the AdS/CFT description.   Specifically, let us study the Mellin amplitude and break it into two sets of terms, those involving $\delta_{si}$ poles and those without such poles.  When all of the $|\delta_{ab}| \gg 1$, a region that is accessible given the large cutoff on dimensions, we can write the Mellin amplitude as
\be
\label{eq:MellinSoftSplit}
M(\delta_{ab}) = F(\delta_{ab}) + \sum_{i=1}^n \frac{R_i(\delta_{ab})}{\delta_{si}} ,
\ee
where the $F(\delta_{ab})$ terms do not have poles in $\delta_{si}$.  Let us assume that the AdS EFT description is sufficiently generic so that the coupling of the spin $\ell$ field dual to the spin $\ell$ CFT operator is not parametrically suppressed\footnote{beyond the usual EFT power-counting $1/(\Delta_\Lambda R^{-1})$ factors.} by factors of order $\Delta_\Lambda$ or larger.  Then when all $|\delta_{ab}| \lesssim \Delta_\Lambda$, the two types of terms in equation (\ref{eq:MellinSoftSplit}) will be of comparable size.  This means that when we take the soft limit with $\delta_{si} \ll \delta_{ij}$, the second set of terms in equation (\ref{eq:MellinSoftSplit}) will dominate, and we can approximate the Mellin amplitude by neglecting the $F(\delta_{ab})$ term, justifying the soft theorems.

The soft approximation becomes arbitrarily good if we can take the limit $\Delta_\Lambda \to \infty$, but only if the coupling of the operator $\CO_\ell$ to the other operators does not vanish in this limit.  This is the flat space limit of AdS/CFT, and so it reproduces the S-Matrix soft theorems if the couplings of the spin $\ell$ particle are non-vanishing in the limit.  For a given (effective) CFT with fixed $\Delta_\Lambda$, the soft theorems will receive corrections of order $\frac{1}{\Delta_\Lambda}$, so if the spin $\ell$ field couples in such a way that its interactions are naturally suppressed by a comparable factor, then the soft theorems will not apply.  The soft theorems could also be invalidated by the presence of an infinite tower of low dimension single-trace operators.  Vasiliev theories \cite{Vasiliev1,Vasiliev2} may have both of these features; it would be interesting to gain a better understanding of their status as effective field theories in AdS.

\subsection{Higher Spins }

\begin{figure}
\begin{center}
\includegraphics[width=\textwidth]{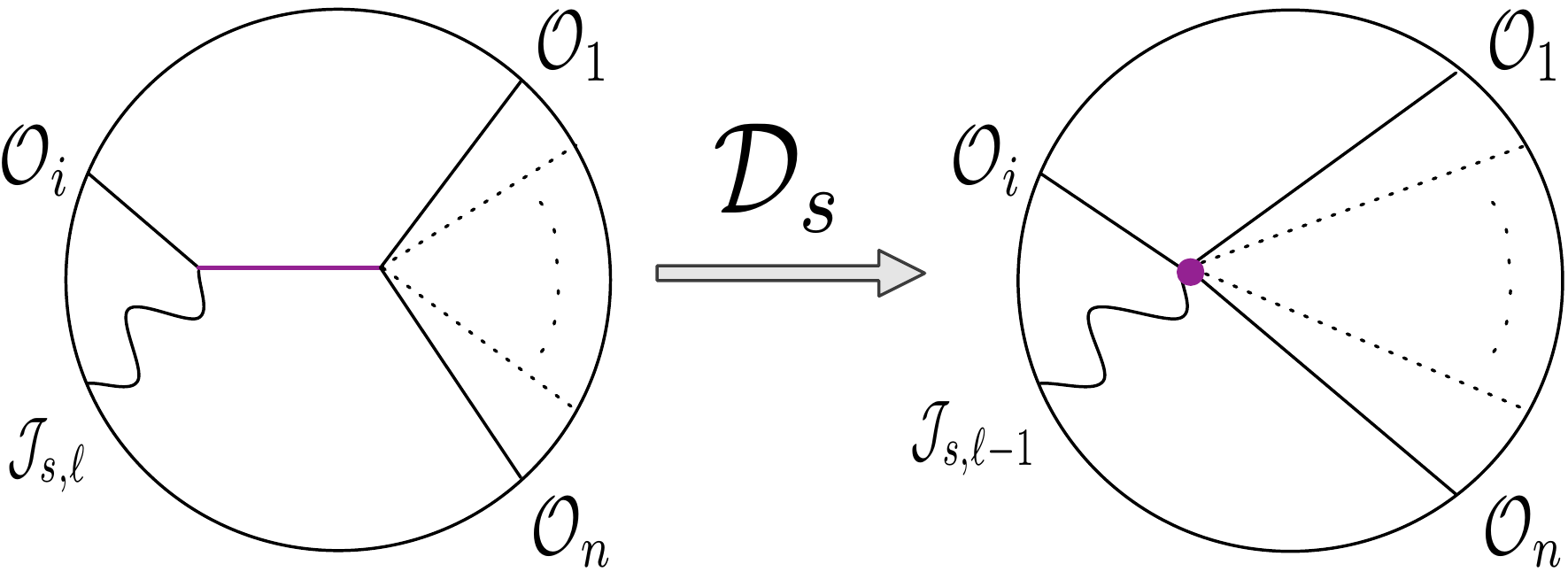}
\caption{This figure shows how imposing current conservation by acting with the differential operator $\mathcal{D}_s$ transforms AdS diagrams relevant in the soft limit.  Their propagators collapse, creating a contact interaction with an effective operator of spin $\ell - 1$.  The left hand diagram is represented by equation (\ref{eq:SpinLDiagram}) while the right diagram corresponds to equation (\ref{eq:SpinLDiagramContact}). This process is formally and conceptually analogous to the S-Matrix manipulations that use Lorentz invariance (or gauge invariance) to obtain equation (\ref{eq:SoftThmConstraint}) from equation (\ref{eq:SoftLimitScattering}) in the soft limit. }
\label{fig:CurrentConservation}
\end{center}
\end{figure}

We have seen how Weinberg's soft theorems can be generalized to an example with vector CFT currents.  Now let us see how to apply the result to higher spins.  For simplicity we will assume that the other operators in the correlator are scalars. This section will make use of some technical results from \cite{NaturalLanguage}, and we refer readers to that reference for details that we omit for the sake of brevity.

First we need to construct the CFT Mellin amplitudes appropriate for the soft limit analysis.  Just as in section \ref{sec:Spin1Soft}, we want to study contributions to the CFT correlator that are enhanced in the limit that $\delta_{si} \ll \delta_{ij}$, and these are the terms with poles at small $\delta_{si}$.  In perturbation theory, these terms are associated with conformal blocks involving the exchange of single-trace operators $\CO_I$ between the product of operators $\mathcal{J}_s \times \CO_i$ and the other $n-1$ operators $\CO_j$ in the $n+1$ point correlator.  With an AdS field theory description, these are just the AdS Feynman diagrams pictured in figure \ref{fig:SoftLimitDiagrams}.   As in section \ref{sec:Spin1Soft}, we can compute these diagrams by acting on their purely scalar analogs with the differential operator $D_{si}^\ell$, which creates the spin of $\CJ_s$.  The validity of this procedure follows from the same logic we used in section \ref{sec:ConformalBlocksandDiagrams}:  we know that it yields Mellin amplitudes with the correct poles and residues, so up to purely polynomially corrections, it must be the correct result. 

The most useful implementation of these ideas represents the diagram of figure \ref{fig:SoftLimitDiagrams} using a form of the AdS propagator \cite{NaturalLanguage} 
\be
G_{BB}(X,Y) = \int_{-i\infty}^{i \infty} \frac{dc}{2 \pi i} \frac{2c^2}{c^2 - (\Delta_i - h)^2}  \int d^d P \frac{\CC_{h+c}}{(-2 P \cdot X)^{h+c} } \frac{\CC_{h-c}}{(-2 P \cdot Y)^{h-c}}.
\ee
Integrating over the bulk points $X$ and $Y$ allows us to express the AdS diagram in a very useful factorized form, in terms of an exchange of an `off-shell' CFT operator $\CO_I$ with dimension $h \pm c$ to the left and right of the propagator.  This gives
\be
A_{i, \ell}^{\mathrm{exch}} (Z_s, P_s; P_j) = \int \frac{dc}{2 \pi i}  \int d^d P \ \! D_{si}^\ell \langle \CO_s(P_s) \CO_i(P_i) \CO_I(P)  \rangle \left( \frac{2c^2}{c^2 - (\Delta_i - h)^2} \right) A(P; P_{j\neq i}) 
\label{eq:SpinLDiagram},
\ee
where $\CO_s$ is a dimension $d - 2 + 2 \ell$ scalar operator that $D_{si}^\ell$ transforms into the conserved current $\mathcal{J}_s$ with dimension $d-2+\ell$.  The intermediate operator $\CO_I$ effectively has the dimension $h+c$, and the factor in parentheses connects the 3-pt and $n$-pt correlators with an AdS scalar bulk-to-bulk propagator.  The pole in $c$ at $\pm (\Delta_i - h)$ sets the dimension of $\CO_I$ to $\Delta_i$, so that `on-shell' we are propagating $\CO_i$, but in this equation $\CO_I$ is the `off-shell' version of $\CO_i$ (in a sense, it is the AdS field to which $\CO_i$ is dual), because it corresponds to an off-shell intermediate propagator in AdS.  Now we need to study the action of the current conservation operator $\mathcal{D}_s$ on this correlator.

In principle we could carry out the analysis in a purely algebraic way, by studying the action of $\mathcal{D}_s$ on $D_{si}^\ell$, as we did for $\ell = 1$ in equation (\ref{eq:ConservationDotDerivative}).  However, it will be much easier technically to study the action of $\mathcal{D}_s$ directly on equation (\ref{eq:SpinLDiagram}) above, because the factor of $[c^2 - (\Delta - h)]^{-1}$ can be directly interpreted as a tree-level propagator in AdS \cite{JoaoMellin, NaturalLanguage}.   Eliminating this factor means erasing the propagator. In other words,
\be
A_{i, \ell-1}^{\mathrm{contact}} (Z_s, P_s; P_j) =  \int \frac{dc}{2 \pi i}  \int d^d P \ \! D_{si}^{\ell-1} \langle \CO_s(P_s) \CO_i(P_i) \CO_I(P)  \rangle \left( 2c^2 \right) A(P; P_{j\neq i}) 
\label{eq:SpinLDiagramContact}
\ee
represents a contact interaction without the AdS propagator.  The relation between equation (\ref{eq:SpinLDiagram}) and equation (\ref{eq:SpinLDiagramContact}) is identical to the relationship between equations (\ref{eq:SoftLimitScattering}) and (\ref{eq:SoftThmConstraint}).    The AdS diagrams and the action of current conservation are pictured in figure \ref{fig:CurrentConservation}.

We will now show that imposing current conservation on equation (\ref{eq:SpinLDiagram}) reduces it to equation (\ref{eq:SpinLDiagramContact}).  First, note that  if we define (with a slight difference from \cite{Costa:2011dw})
\be
V_{si} = (Z_s \cdot P_i)(P_s \cdot P) - (Z_s \cdot P)(P_s \cdot P_i),
\ee
then the action of $D_{si}^\ell$ on equation (\ref{eq:SpinLDiagram}) can be straightforwardly computed to be 
\be
A_{i, \ell}^{\mathrm{exch}} (Z_s, P_s; P_j) &=& \int \frac{dc}{2 \pi i}  \int d^d P \ \! \left(c + (\Delta_i - h) - \ell + 1 \right)_{\ell} \left( \frac{V_{si}}{P \cdot P_i} \right)^\ell \langle \CO_s(P_s) \CO_i(P_i) \CO_I(P)  \rangle 
\nn \\
&& \times \left( \frac{2c^2}{c^2 - (\Delta_i - h)^2} \right) A(P; P_{j\neq i}) ,
\ee
where we recall that the Pochhammer symbol $(a)_b = \Gamma(a+b)/\Gamma(a)$.  This equation can be derived noting that $D_{si}$ only differentiates with respect to $P_i$, and $P_i$ only appears in the 3-pt correlator of $\CO_s$, $\CO_i$, and $\CO_I$, which is uniquely determined by conformal symmetry.  

Now we need to apply the current conservation operator $\mathcal{D}_s$ \cite{Costa:2011dw} defined in equation (\ref{eq:CurrentConsOp}).  A straightforward computation shows that
\be
[ \mathcal{D}_s, V_{si}^\ell ] &=& (h-1+\ell) \ell (\ell-1) V_{si}^{\ell-2} (P_i \cdot P) ( (Z_s \cdot P_i)(P_s \cdot P) + (Z_s \cdot P)(P_s \cdot P_i)) \nn \\
&& + \ell V_{si}^{\ell-2} \left((h-2+\ell) V_{si} V_Z^A + (\ell-1) (P_s \cdot P_i) (P_s \cdot P) (P_i \cdot P) Z_s^A  \right) \nabla_{s A}.
\ee
Using this we can now calculate that 
\be
\mathcal{D}_s A_{i, \ell}^{\mathrm{exch}} (Z_s, P_s; P_j) &=& \int \frac{dc}{2 \pi i}  \int d^d P \ \! \left(c + (\Delta_i - h) - \ell + 1 \right)_{\ell-1} \left( \frac{V_{si}}{P \cdot P_i } \right)^{\ell-1}  \ell (h-2+\ell)
\nn \\
&& \times \langle \CO_s(P_s) \CO_i(P_i) \CO_I(P)  \rangle  \left(c^2 - (\Delta_i - h)^2 \right) \left( \frac{2c^2}{c^2 - (\Delta_i - h)^2}  \right) A(P; P_{j\neq i}) \nn \\
&=&  \ell (h-2+\ell) A_{i, \ell-1}^{\mathrm{contact}} (Z_s, P_s; P_j) 
\label{eq:DAexchEqualsAcont},
\ee
as desired.    This followed from applying current conservation to a single diagram, where $\CJ_s$ couples to the operator $\CO_i$.  To obtain this result, it was crucial to use the fact that the dimension of $\CJ_s$ is $d-2+\ell$, otherwise there are additional terms that do not have a simple interpretation as a contact interaction in AdS.  To use this to obtain an AdS/CFT soft theorem we need to sum over diagrams; the current will be conserved only if
\be
0 = \sum_{i=1}^n  A_{i, \ell-1}^{\mathrm{contact}} (Z_s, P_s; P_j),
\ee
where the relative couplings $g_i$ of $\CJ_s$ to $\CO_i$, which can be precisely defined in terms of the magnitude of the three point function of these operators with $\CO_i^\dag$, have been absorbed implicitly into $A_{i,\ell-1}^{\rm contact}$.  In the case where $\ell = 1$ and $n=4$ we already saw that all of the $A_{i, 0}^{\mathrm{contact}} $ terms are identical up to $g_i$, so we must have 
\be
\sum_{i=1}^n g_i = 0
\ee
or charge conservation for the $\ell = 1$ current. We show using the methods of \cite{NaturalLanguage} in appendix \ref{sec:ComputationContactTerm} that the Mellin amplitude for the contact terms is approximately universal in the soft limit, so that
\be
A_{i, \ell-1}^{\mathrm{contact}} (Z_s, P_s; P_j) \approx D_{si}^{\ell-1} \int [d \delta] g_i M_{s}(\delta_{ab}) \prod_{a<b}^{n+1} \Gamma(\delta_{ab}) P_{ab}^{-\delta_{ab}}.
\ee
 $M_s$ is the universal Mellin amplitude, we have included $P_s$ as the $n+1$ coordinate, and the equation only holds in the soft limit regime, with $\delta_{si} \ll \delta_{ij}$.  Now we see that the details of the Mellin amplitude $M_s$ factor out, and the current conservation condition becomes
\be
\sum_{i=1}^n g_i D_{si}^{\ell-1} \prod_{a<b}^{n+1} P_{ab}^{-\delta_{ab}} = 0.
\ee
Again, in the case $\ell = 1$, we find that the sum of the charges must vanish.  When $\ell = 2$, the case of gravity, we can compute the derivative to find
\be
0 &=& \sum_{i=1}^n g_i D_{si} \prod_{a<b}^{n+1} P_{ab}^{-\delta_{ab}} \nn \\
&=& \sum_{i \neq j}^n g_i \delta_{ij} \left[\frac{(Z_s \cdot P_j) (P_s \cdot P_i) - (Z_s \cdot P_i) (P_s \cdot P_j)}{P_i \cdot P_j} \right] \prod_{a<b}^{n+1} P_{ab}^{-\delta_{ab}}.
\ee
For generic $P_i$ and $\delta_{ij}$, this condition can be satisfied only if the charges $g_i$ are universal, so that $g_i = G$, in which case the formula on the last line vanishes due to the symmetry of $\delta_{ij}$ and the anti-symmetry of the expression in square brackets.  Thus we have shown that a conserved spin $2$ tensor must couple universally in the AdS/CFT soft limit.  Higher spins with $\ell > 2$ produce more complicated tensor structures in $Z_s$, $P_s$, and $P_i$ with $\delta_{ij}$ coefficients, and the sum of these structures cannot vanish without imposing kinematic constraints on the CFT correlation function.  So AdS field theories with a finite number of fields, generic couplings, and a cutoff that is large in AdS units cannot experience long-range forces from higher-spin particles, as expected from flat space field theory.

\section{Discussion}
\label{sec:Discussion}

Recent work has shown that Mellin space provides an extremely natural language for discussing CFT correlation functions dual to AdS quantum field theories \cite{JoaoMellin, NaturalLanguage, Paulos:2011ie}.  Here we have argued the converse:  that by looking at CFT correlators in Mellin space, one can straightforwardly deduce whether or not they can arise from an effective field theory in AdS.  

We formulated three criteria that are necessary and sufficient for an AdS field theory description of a given CFT.  The first two criteria require a perturbative `$1/N$' expansion of the CFT correlators and Hilbert space of states about a conformal Fock space generated by a finite set of `single-trace' operators with Gaussian correlators.  This means that there must be a gap in the CFT spectrum between a finite set of low-dimension single-trace operators and the other single-trace operators.  These two criteria are fairly well-known and have been thoroughly discussed \cite{JP}.  However, by themselves they are insufficient to guarantee the existence of an AdS effective field theory description.

We have introduced a third criterion, that the CFT correlators are polynomially bounded in Mellin space, or equivalently, that they can be well-approximated by the first few terms in a Mellin-space power series expansion identical to familiar EFT expansions in energy.  In previous work \cite{JP, Heemskerk:2010ty} conditions far more restrictive than our third criterion were required for the technical analysis.  Their conditions \cite{JP} imply a purely polynomial Mellin amplitude, as mentioned in \cite{JoaoMellin}, but without Mellin space technology, their analysis was very involved and, more importantly, it was very difficult to see how to weaken and generalize their assumptions.  

We have not shown how to begin with an abstract, top-down definition of a CFT (for example in terms of a gauge theory Lagrangian) and determine if it has an AdS field theory description.  But we believe that we have largely solved the problem from the bottom up: given the CFT correlation functions themselves, we have explained how to determine if a dual AdS field theory exists, and how to understand bulk locality at length scales far below the AdS length $R$.  As we have emphasized elsewhere \cite{Analyticity, Unitarity}, the flat space limit of AdS/CFT provides an elegant method for studying sub-AdS scale locality via scattering theory, and we reviewed in section \ref{sec:FSLandAnalyticity} how the Mellin amplitude defines an S-Matrix with analyticity properties that follow from bulk locality.  

There remains the possibility that most complete unitary CFTs obeying our first two criteria automatically satisfy our third criterion.  In section \ref{sec:Unitarity} we argued that considerations of unitarity support this conclusion, in the sense that Mellin amplitudes that grow in an uncontrolled way at large $\delta_{ij}$ generically violate perturbative unitarity at a low scale. The integrability program \cite{Beisert:2010jr} has shown that suitable generalizations of our first two criteria to AdS$_5 \times S^5$ hold at large 't Hooft coupling in $\mathcal{N}=4$ SYM.   It would be interesting to study perturbative unitarity constraints in more detail, because if one could also show that our third criterion is obeyed by the $\mathcal{N}=4$ theory at large 't Hooft coupling, our results might eventually be combined with integrability to produce a reasonably rigorous proof of the AdS/CFT correspondence in this case.  

In section \ref{sec:HigherSpin} we made a modest show of generalizing Weinberg's soft theorems for the S-Matrix to AdS/CFT correlation functions.  Our goal here was to give at least a partial explanation of why CFTs with AdS field theory descriptions tend not to have low-dimension single-trace operators of high-spin.  There are well-known difficulties (see e.g. \cite{Porrati:2008rm}) in formulating field theories with light, high-spin particles, so physically it is already clear why low-dimension high-spin operators may be problematic.  In section \ref{sec:HigherSpin} we made this argument  much more precise by showing how Weinberg's argument can be applied to AdS/CFT using Mellin space technology.  We found another set of striking similarities between Mellin space and momentum space, even at the level of the technical steps in the argument.  However, unlike the S-Matrix in flat space, AdS effectively has both a UV and an IR cutoff, and the `soft limit' cannot be taken below the AdS scale.  This leaves loopholes for AdS field theories that are unavailable in flat space.  It would be very interesting to better understand how Vasiliev theories \cite{Vasiliev:1990en, Vasiliev1,Vasiliev2} exploit these loopholes. 

\section*{Acknowledgments}

We would like to thank Ami Katz,  Steve Shenker, Raman Sundrum, and especially Jo\~ ao Penedones for discussions.  We would also like to thank the organizers of the Back to the Bootstrap 2 workshop at the Perimeter Institute for their hospitality, and the participants at that conference for many useful conversations and lessons.  This material is based upon work supported in part by the National Science Foundation Grant No. 1066293 and the hospitality of the Aspen Center for Physics.   ALF was partially supported by ERC grant BSMOXFORD no. 228169. JK acknowledges support from the US DOE under contract no. DE-AC02-76SF00515.

\appendix

\section{Computation of General Spin Contact Term}
\label{sec:ComputationContactTerm}

In this appendix we will compute the contact terms that arise after imposing current conservation on a given soft-limit AdS Feynman diagram, namely
\be
A_{i, \ell-1}^{\mathrm{contact}} (Z_s, P_s; P_j) = \int \frac{dc}{2 \pi i}  \int d^d P \ \! D_{si}^{\ell-1} \langle \CO_s(P_s) \CO_i(P_i) \CO_I(P)  \rangle \left( 2c^2 \right) A(P; P_{j\neq i}) ,
\ee
which is equation (\ref{eq:SpinLDiagramContact}) in the body of the paper.  We will show that for all $i$, it takes the universal form
\be
A_{i, \ell-1}^{\mathrm{contact}} (Z_s, P_s; P_j) \approx D_{si}^{\ell-1} \int [d \delta] g_i M_{s}(\delta_{ab}) \prod_{a<b}^{n+1} \Gamma(\delta_{ab}) P_{ab}^{-\delta_{ab}}
\ee
in the soft limit, where $\delta_{si} \ll \delta_{ij}$, and $M_s$ is a universal Mellin amplitude, independent of $i$.  

To perform the computation, we can use the results of \cite{NaturalLanguage},  specifically their equations (53), (54), and (61).  This allows us to write the Mellin amplitude for $A_{i, \ell-1}^{\mathrm{contact}}$ without the $D_{si}^{\ell-1}$ in a form that will be amenable to making the approximation that $\delta_{is} \ll \delta_{ij}$.  So, take the scalar correlator $S_{s, i, \ell-1}^{(\rm contact)}$ to be $A_{i,\ell-1}^{\rm (contact)}$ without the action of $D_{si}^{\ell-1}$:
\be
&&S_{s,i, \ell-1}^{\mathrm{(contact)}} (P_s; P_j) \equiv \int \frac{dc}{2 \pi i}  \int d^d P \ \! \langle \CO_s(P_s) \CO_i(P_i) \CO_I(P)  \rangle \left( 2c^2 \right) A(P; P_{j\neq i}) \\
&& = \left[ \frac{1}{2} \left( \mathcal{L}_s + \mathcal{L}_i  \right)^2  - \Delta_i (d - \Delta_i) \right]  
\int \frac{dc}{2 \pi i}  \int d^d P \ \!  \langle \CO_s(P_s) \CO_i(P_i) \CO_I(P)  \rangle \left( \frac{2c^2}{c^2 - (\Delta_i - h)^2} \right) A(P; P_{j\neq i}). \nn
\ee
  The integral is just a bulk-to-bulk propagator connecting a scalar 3-point amplitude to a scalar $n$-point amplitude,
of the form of equation (53) in \cite{NaturalLanguage}, and thus applying the factorization formula equation (43) in \cite{NaturalLanguage} determines its poles and residues in $\delta_{is}$:
\be
\CI &=& \sum_m \frac{-4 \pi^h \Gamma^2(\Delta_i - h + 1)}{\Gamma(\Delta_i -h+1+m) (\Delta_s -2m -2 \delta_{is})} M_L \frac{1}{m! (\frac{\Delta_s}{2})_{-m}} R_m(\delta_{jk}).
\ee
The factor $M_L$ is just the Mellin amplitude for the three-point function $\langle \CO_s(P_s) \CO_i(P_i) \CO_I(P)  \rangle$, and so is just a constant, and the factor $R_m(\delta_{jk})$ is related to the Mellin amplitude $M_R(\delta_{jk})$ for the $n$-point function $A(P; P_{j\neq i})$ according to \cite{NaturalLanguage}
\be
R_m(\delta_{jk}) &=& \left. \sum_{\Sigma n_{jk} = m} M_R(\delta_{jk} + n_{jk}) \prod_{{j<k \atop j,k \ne i}} \frac{(\delta_{jk})_{n_{jk}}}{n_{jk}!}\right|_{\delta_{is} = \frac{\Delta_s}{2} -m} .
\ee
Now, the key point is that the hard variables $\delta_{ij}$ are all much larger than $\delta_{is} \gtrsim 1$ by assumption.  Furthermore, since $\Delta_s = 2 (h-1+\ell)$ is a $\CO(1)$ number, the prefactor
$\frac{1}{m! (\frac{\Delta_s}{2})_{-m}} $ causes an exponential shutdown of the residues of poles at large $m$.  Thus, the shifts of the $\delta_{ij}$ in $R_m(\delta_{ij})$ from the $n_{ij}$ (which are all $\le m$) and in their constraints from having to satisfy $\sum_{i\ne j} \delta_{ij} = \Delta_i$ are all completely negligible in this limit.  Thus $M_R$ factors out, and we can approximate
\be
R_m(\delta_{ij}) &\approx& M_R(\delta_{jk}) \left. \sum_{\Sigma n_{jk} = m}  \prod_{{j<k \atop j,k \ne i}} \frac{(\delta_{jk})_{n_{jk}}}{n_{jk}!}\right|_{\delta_{is} = \frac{\Delta_s}{2} -m} \nn\\
 &=& M_R(\delta_{jk}) \left. \frac{\left(\sum_{{j<k \atop j,k \ne i}} \delta_{jk}\right)_m}{m!} \right|_{\sum_{{j<k \atop j,k \ne i}}\delta_{jk} = \frac{\sum_{k\ne i} \Delta_k - \Delta_i}{2} -m} \nn\\
 &=&M_R(\delta_{jk}) \frac{1}{m! (\frac{\sum_{k\ne i} \Delta_k - \Delta_i}{2})_{-m}}.
\ee
Thus $M_L$ and $M_R$ approximately factor out of the integral, and we find
\begin{equation}
\CI \approx M_L M_R(\delta_{jk}) \sum_m \frac{-4 \pi^h \Gamma^2(\Delta_i - h + 1)}{\Gamma(\Delta_i -h+1+m) (\Delta_s -2m -2 \delta_{is})} \frac{1}{m! (\frac{\Delta_s}{2})_{-m}}  \frac{1}{m! (\frac{\sum_{k\ne i} \Delta_k - \Delta_i}{2})_{-m}} .
\end{equation}
Now, recall that $S_{s,i, \ell-1}^{\mathrm{(contact)}} (P_s; P_j) = \left[ \frac{1}{2} \left( \mathcal{L}_s + \mathcal{L}_i  \right)^2  - \Delta_i (d - \Delta_i) \right] \CI$.  Since the conformal Casimir involves shifts of the Mellin variables only by small integers, $M_R(\delta_{jk})$ is effectively constant, and to good approximation the Casimir acts only on the poles in the sum above. Therefore $\left[ \frac{1}{2} \left( \mathcal{L}_s + \mathcal{L}_i  \right)^2  - \Delta_i (d - \Delta_i) \right] $ just produces the constant $\lambda_{n+1}/(\lambda_n \lambda_3)$, where $\lambda_n$ was defined in equation (\ref{eq:lambdan}) for an $n$-point contact Mellin amplitude and depends only on the dimension of the external operators.  Here, $\lambda_{n+1}$ is for dimensions $\{\Delta_i\}_{i=1,\dots,n}$ and $\Delta_s$, $\lambda_n$ is for dimensions $\{\Delta_i\}_{i=1,\dots,n}$, and $\lambda_3$ is for $(\Delta_s, \Delta_i, \Delta_i)$.  We therefore obtain
\be
 S_{s,i, \ell-1}^{\mathrm{(contact)}} (P_s; P_j) \approx \frac{\lambda_{n+1} }{\lambda_n} \frac{M_L}{\lambda_3} M_R(\delta_{jk}) \equiv g_i M_s ,
\ee
with $g_i \equiv \frac{M_L}{\lambda_3}$. Thus, our universal Mellin amplitude $M_s$ is manifestly independent of the leg (operator $\CO_i$) the current is attached to, as claimed.

\section{Comments on an Argument of Sundrum}
\label{sec:Sundrum}

Sundrum \cite{Sundrum:2011ic} has made a rather different (and more obviously physical) argument for AdS locality that nevertheless contends with issues related to our third criterion.  He suggests that conformal invariance allows us to `rotate' into the extra dimension in order to guarantee bulk locality from CFT locality.  We believe his argument implies AdS causality but that it fails to justify microscopic AdS locality because, roughly speaking, he is missing our third criterion.  

To make the argument in detail, he studies a conformally invariant large $N$ gauge theory in 4-dimensional Minkowski space, and introduces an IR cutoff $\Lambda_Q$ interpreted as the dual of an IR brane in the AdS Poincar\' e patch.  With conformal invariance broken, we have an S-Matrix that describes the scattering of the various glueball states with masses $m_i \sim \Lambda_Q$.  The Large $N$ expansion implies that to leading order, the 2-to-2 scattering amplitudes of glueballs can be written as
\be
\mathcal{A} = \frac{1}{N^2} \left[ Q(s,t, \Lambda_Q) + \sum_i k_i  \Lambda_Q^2 \frac{P_{\ell_i} (\cos \theta)}{s - m_i^2} \right],
\ee
where the sum is over the various glueballs in the theory.   The $k_i$ are order one coefficients, 
 $P_{\ell_i}$ are Legendre polynomials associated with the spins, and the term $Q(s,t,\Lambda_Q)$ is a contact interaction piece that is generically an infinite power series in $s$ and $t$.

Sundrum invokes large $N$ gauge theory lore to argue that at least at this order in $1/N$, the S-Matrix is local \cite{Sundrum:2011ic}. Then he proceeds to argue that this implies AdS locality at leading non-trivial order in $1/N$.  What he means by this is that there exists something like a local bulk string theory or string field theory, in the sense that there are an infinite number of modes with interactions that look like local polynomials in fields and derivatives.  We do not disagree with this interesting line of reasoning, but by itself, it only produces an extremely convoluted AdS `field theory' with a cutoff around the AdS scale, $1/R$.  In other words, this part of the argument suggests AdS causality, but not microscopic locality.

However, Sundrum then invokes the existence of a gap in the gauge theory spectrum to argue for a larger cutoff on the AdS EFT.   We do not agree that the existence of a gap demonstrably lifts the cutoff; if it did then our third criterion would not be necessary.  Even if the contributions of operators above the gap are suppressed, this does not automatically lift the AdS EFT cutoff on the interactions of the low-dimension operators (light fields in AdS), which naively could remain at the AdS scale.  We gave an argument in section \ref{sec:Unitarity} that in some sense one can generically expect that the AdS EFT cutoff lifts with the gap, but our argument falls far short of a proof, and these subtleties are left unexamined in \cite{Sundrum:2011ic}.  To state it once more:  the existence of a perturbative $1/N$ expansion in the bulk does not automatically guarantee an EFT-type expansion with a cutoff at parametrically large energy.

We should add that Sundrum could use a more expansive definition of locality.  For example, the perturbative glueball S-Matrix is exactly analytic.  It is not entirely clear to us what this should imply for the formulation of the bulk theory, although we can point to a vaguely analogous line of reasoning from our point of view.  A great way to study microphysical locality in AdS is via bulk scattering, and we know that the S-Matrix of the bulk theory can be obtained from a flat space limit of AdS/CFT in terms of the Mellin amplitude \cite{JoaoMellin, NaturalLanguage, Analyticity, Unitarity}.  All CFTs give rise to meromorphic Mellin amplitudes with only simple poles, and large $N$ CFTs have correlators that are just a $1/N$ perturbation away from free propagation in AdS.  As we mentioned in \cite{Analyticity}, this suggests that one can formally compute an S-Matrix for these theories (although it may vanish).  One can hope that it will inherit fairly nice analyticity properties from the Melin amplitude, and that this could provide a reasonable, albeit very general definition of bulk locality.

\bibliographystyle{utphys}
\bibliography{BoundednessBib}

 \end{document}